\documentclass{aastex63}

\received{\today}
\revised{}
\accepted{}

\submitjournal{ApJ}

\shorttitle{AGN Outflows}
\shortauthors{Tanner}

\usepackage{amsmath,amssymb,amstext,xspace}

\newcommand{\kms}[0]{km~s$^{-1}$\xspace}
\newcommand{\msun}[0]{M$_{\Sun}$\xspace}

\newcommand\ergs{erg~s$^{-1}$\xspace}

\newcommand{\hotsx}{Hot$_{sX}$\xspace}
\newcommand{\hotmx}{Hot$_{mX}$\xspace}
\newcommand{\hothx}{Hot$_{hX}$\xspace}

\begin{document}

\title{Simulations of AGN Driven Galactic Outflow Morphology and Content}

\correspondingauthor{Ryan Tanner}
\email{ryan.tanner@nasa.gov}
\author[0000-0002-1359-1626]{Ryan Tanner}
\affiliation{NASA Goddard Space Flight Center\\
Greenbelt, Maryland, 20771, USA}
\author{Kimberly A. Weaver}
\affiliation{NASA Goddard Space Flight Center\\
Greenbelt, Maryland, 20771, USA}

\begin{abstract}
Using a series of 3D relativistic hydrodynamical simulations of active galactic nuclei (AGN) we investigate how AGN power, a clumpy ISM structure, and AGN jet angle with respect to the galactic disk affect the morphology and content of the resulting galactic outflow. For low power AGN across three orders of magnitude of AGN luminosities ($10^{41}-10^{43}$ \ergs) our simulations did not show significant changes to either the morphology or total mass of the outflow. Changing the angle of the AGN jet with respect to the galaxy did show small changes in the total outflow mass of a factor of 2-3. Jets perpendicular to the galactic disk were created hot single phase outflows, while jets close to parallel with the disk created the outflows were multi-phase with equal parts warm and hot, and significant cold gas. Overall The final morphology of low power AGN outflows depends primarily on how the jet impacts and interacts with large, dense clouds in the clumpy ISM. These clouds can disrupt, deflect, split, or suppress the jet preventing it from leaving the galactic disk as a coherent structure. But for simulations with AGN luminosities $> 10^{44}$ \ergs the ISM played a minor role in determining the morphology of the outflow with an undisrupted jet leaving the disk. The final morphology of AGN outflows is different for low power AGNs vs. high power AGNs with the final morphology of low power AGN outflows dependent on the ISM structure within the first kpc surrounding the AGN.
\end{abstract}


\section{Introduction}\label{sec:intro}
Active Galactic Nuclei (AGN) are important drivers of galactic outflows and key components of galactic evolution. 
Astronomers are confident that AGN-driven outflows are generated by radiation pressure and mechanical energy from jets created around the central supermassive black hole
\citep{1989ApJ...345L..21B,2011ApJ...740..100B,2016A&ARv..24...10T,2018NatAs...2..198H}.
The shape of the extended radio or X-ray emission from AGN-driven outflows is now generally agreed to be highly influenced by the local ISM on various spatial scales 
\citep{2011ApJ...740..100B, 2016A&ARv..24...10T, 2019ARA&A..57..467B,  2020NewAR..8801539H, 2020ApJ...896...18P}, 
although there are suggested links between the AGN accretion mode and jet morphology 
\citep{10.1111/j.1365-2966.2012.20414.x,2016A&ARv..24...10T,2018NatAs...2..273H}.
The diagnostic power of AGN-driven outflows has materialized in an ever-growing body of work using outflow properties to test AGN feedback \citep{2018NatAs...2..198H}.

The fact that AGN-driven outflows possess a broad range of morphologies \citep[e.g.,][]{2020NewAR..8801539H} presents a challenge for theorists. The content of AGN jets is still not well constrained, and models generally assume either an electron-position plasma, or an electron-proton plasma
\citep{2009A&ARv..17....1W,2019ARA&A..57..467B}.
Along with the jet, multiple gas phases will be entrained in the outflow
\citep{2012ApJ...760...77A,2013Sci...341.1082M,2013ApJ...768...75R,2013ApJ...776...27V,2014A&A...562A..21C,2014MNRAS.441.3306H,2017A&A...601A.143F,2018NatAs...2..198H,2019A&A...627A..53H,2020A&A...635A..47H,2020MNRAS.497.5229C,2021A&A...645A.130C}.
Outside of the galaxy, warm and cool gas will also condense out of the hot halo when perturbed by shocks from the outflow \citep{2018ApJ...854..167G}. 
The plasma in the jet produces observable radio emission through synchrotron radiation
\citep[For a review see][]{2019ARA&A..57..467B}, while X-rays trace shocks and the hot outflowing gas \citep{10.1093/mnras/168.3.479,hardcastle2000environments,2002NewAR..46..365B,Birzan_2004,2009A&ARv..17....1W,doi:https://doi.org/10.1002/9783527641741.ch7,2020NewAR..8801539H}. 
Observations have shown that most radio jets are surrounded by a cocoon, or shell of hot X-ray emitting gas \citep{2017ApJ...851...87O,2018A&A...619A..75M,2018ApJS..235...32S,2020MNRAS.497..482L,2020ApJ...894..157M,2021ApJS..252...31J,2021A&ARv..29....3O}. 

The study of radio emission from active galaxies has a vast history. 
The emission can be classified based on the size and the shape of the SED \citep{2012ApJ...760...77A,2020ApJ...896...18P,2021A&ARv..29....3O}. 
At the smallest spatial scales compact symmetric objects (CSOs) appear to represent the beginning of the evolution of AGN driven outflows, with GHz-peaked spectrum (GPS) sources and compact steep-spectrum (CSS) sources representing their later evolution. 
At the largest spatial scales radio lobes can be classified as either Fanaroff–Riley (FR) type I or FR type II \citep{1974MNRAS.167P..31F}. FR I radio lobes appear as more compact with less well defined jets. 
The outer lobes of FR I jets are also dimmer than immediately around the AGN. FR II radio lobes have longer, more well defined jets ending in bright ``hot spots" that make up more than half of the total radio luminosity. 
A simple description would be that FR I are ``center brightened", while FR II are ``edge brightened" \citep{2019MNRAS.488.2701M}. 

Is is not clear whether the ultimate morphology of the outflows are due to effects of the $\approx 1$ kpc scale ISM structure of the host galaxy 
\citep{1993ApJ...405L..13D,1995ApJS..101...29B,1997MNRAS.286..215K,2000A&A...363..507G,2006A&A...455..161L,2019MNRAS.488.2701M}
or to interactions with small scale structures ($< 1$ pc) and the environment around the supermassive black hole
\citep{rees1971new,2007MNRAS.376.1849H, 2008ApJ...685..147N, 2008ApJ...685..160N, 2014MNRAS.440..269M}. 
Surveys of FR I and II galaxies indicate that radio lobe morphology is independent of AGN accretion modes and instead dependent on the jet's interaction with the larger environment
\citep{2013MNRAS.430.3086G,2019MNRAS.489.2177A,2019MNRAS.488.2701M,2020ApJ...905...42G}. But the specifics of these interactions on the lobes is still not well understood as this extends to objects with weaker and much less well defined outflows.

Previous theoretical work has been motivated toward examining jets and cocoons across a range of AGN powers \citep{2009A&ARv..17....1W, 2012ApJ...757..136W, 2013ApJ...772L...1M, 2016A&A...596A..12M, 2017MNRAS.472.4707B, 2017MNRAS.470.4530W, 2018MNRAS.481.2878E, 2018MNRAS.479.5544M, 2019A&A...621A.132M, 2021NewAR..9201610K, 2021MNRAS.508.5239Y}. 
Our work expands on this by modeling jets in a very non-homogeneous environment. 
For this paper we investigate how a clumpy ISM, on a $\sim$1 kpc scale, affects the shape, kinematics, and content of such an AGN driven galactic outflow. 
We explore a full range of jet inclination angles with respect to the galaxy disk to understand ISM interactions for very highly inclined AGN.
We ran a series of 3D relativistic hydrodynamical simulations with the direction of the jet ranging from perpendicular to the galactic disk to parallel to the disk. 
Because of the clumpy ISM each jet angle impacts on a unique part of the ISM creating a set of different outflow morphologies. 
Our clumpy ISM allows for complex outflow morphologies, such as split flows, deflected jets, and asymmetric outflows. 
Similar simulations have been run in 2D \citep{2005MNRAS.359..781S}, in 3D with a non-uniform ISM and the AGN power near the historical dividing line between FR I and FR II lobes \citep{2007ApJS..173...37S,2009AN....330..287J,MNRAS2011.411..155G}, in 2D with a clumpy medium out to 60 kpc and a high power jet \citep{2009MNRAS.396...61T}, in 3D with a clumpy ISM and high power ($>10^{43}$ \ergs) AGN \citep{2011ApJ...728...29W,2012ApJ...757..136W,2016MNRAS.461..967M,MNRAS2012.425..438G}, and in 2D with high resolution \citep{2020MNRAS.498.3870M}. 
All of the above mentioned simulations are on the level of the local galactic environment comprising the ISM out to a few kpc. 
There are also numerous simulations at a much larger scale ($> 100$ kpc) examining jets in a cluster environment. 
These include hydrodynamic simulations in 2D and 3D \citep{2005A&A...431...45K,2016MNRAS.458..558D,2019MNRAS.490.5807E, 2021MNRAS.502..423S,2021ApJ...920..144S,2021MNRAS.508.5239Y}, magnetohydrodynamic simulations in 2D and 3D \citep{2012ApJ...750..166M,2013MNRAS.430..174H,2014MNRAS.443.1482H,2017MNRAS.470.4530W}, and semi-analytic models based on simulations \citep{2015ApJ...806...59T,2018MNRAS.475.2768H}. 
Our simulations are commensurate with the former simulations of jets interacting with the local ISM and not the larger cluster medium. 
For a review of simulations of jets see \citet{2019Galax...7...24M} or \citet{2021NewAR..9201610K}.

To test how a clumpy ISM affects the morphology of the outflow, we run a set of simulations ranging over AGN inclination angles of $0^{\circ}$ to $90^{\circ}$ with $P_{AGN} = 10^{42}$ \ergs, and then a smaller set with angles of $30^{\circ}$ to $60^{\circ}$ with $P_{AGN} = 10^{41}$ and $10^{43}$\ergs. 
To investigate how the AGN power relates to the break between FR I and II we ran a set of simulations with the AGN power ranging from $10^{41}$ \ergs to $10^{46}$ \ergs to see how AGN power may affect the outflow morphology.
In the radio, the traditional break between FR I and II is $L_{150} \sim 10^{26} W Hz^{-1}$ \citep{FanaroffRiley,1996AJ....112....9L,2012ApJ...760...77A,2019MNRAS.488.2701M}. 
This corresponds to an AGN jet power of $\sim 10^{44.5}$ \ergs \citep{1999MNRAS.309.1017W,2017MNRAS.467.1586I}, which is spanned by the power ranges in or simulations. 
Our full set of simulations with corresponding AGN inclination angles and powers is given in Table \ref{tab:AGNPower}. 

In Section \ref{sec:setup} we explain our setup and how we generate the clumpy ISM. 
We use the same initial conditions for all simulations. In Section \ref{sec:angle} we discuss how the direction of the AGN jet affects the overall morphology of the outflow. 
In Section \ref{sec:AGNpower} we discuss how the AGN jet power influences the morphology of the outflow. 
Then in Section \ref{sec:jetbub} we describe how the outflow consists of two components, a relativistic jet with laminar flow, and a turbulent semi-spherical bubble of hot gas. 
For $P_{AGN} < 10^{44}$ \ergs the evolution of the jet and bubble components depends strongly on the interaction of the AGN jet with the ISM within the first kiloparsec. For $P_{AGN} > 10^{44}$ \ergs the outflow is dominated by the relativistic jet with laminar flow.

\section{Simulation Setup}\label{sec:setup}

We model an AGN driven jet using the public Athena code \citep{Stone-Athena}. 
Because the AGN jet has a relativistic velocity we use the relativistic unsplit MUSCL-Hancock integrator with the relativistic HLLC Riemann solver \citep{2005MNRAS.364..126M} included with Athena. 
The code has been modified \citep{2016ApJ...821....7T} to include radiative cooling down to 10 K, using a cooling function that combines the function given by \cite{Koyama} with tabulated solar metallicity data from \cite{1993ApJS...88..253S}.

We set up a computational space of $5000\times5000\times8000$ pc divided into a grid of $500\times500\times800$ cells giving us a resolution of 10 pc per cell.

In our simulations we tilt the direction of the AGN jet with respect to the vertical (minor) axis of the galaxy. 
Possible values range from between $0^{\circ}$ (perpendicular to the disk) to $90^{\circ}$ (parallel to the disk). 
We also vary AGN luminosities over six orders of magnitude from $10^{41}$ \ergs, to $10^{46}$ \ergs. 
The combinations of AGN luminosities and tilt angles we use are given in Table \ref{tab:AGNPower} for a total of 19 simulations. 
Each simulation is run for a total of 600 kyr at which point in all but two simulations the jets have reached the edge of the computational grid.

\begin{deluxetable}{cD}\label{tab:AGNPower}
\tablehead{
\colhead{AGN Power} & & \\ $(erg~s^{-1})$ &  \multicolumn2c{Angles} 
}
\startdata
$10^{41}$ & $30^{\circ}$, $45^{\circ}$, $60^{\circ}$\\
$10^{42}$ & $0^{\circ}$, $15^{\circ}$, $30^{\circ}$, $45^{\circ}$\\
          & $60^{\circ}$, $75^{\circ}$, $90^{\circ}$\\
$10^{43}$ & $30^{\circ}$, $45^{\circ}$, $60^{\circ}$\\
$10^{44}$ & $30^{\circ}$, $45^{\circ}$\\
$10^{45}$ & $30^{\circ}$, $45^{\circ}$\\
$10^{46}$ & $30^{\circ}$, $45^{\circ}$\\
\enddata
\caption{AGN powers and tilt angles used in simulations.}
\end{deluxetable}

\subsection{Initial Galactic Disk}\label{sec:setup:disk}

We model the gravitational potential with two parts consisting of a spherical stellar bulge and a stellar disk. 
Because we are modeling only the central few kpc we chose not to include the dark matter halo as at this range it would not significantly affect the dynamics of the gas.

The spheroidal bulge $\Phi_{\rm ss}(R)$ is modeled as a King model, 
\begin{equation}\label{eq:phiss}
\Phi_{\rm ss}(R)=-\frac{GM_{\rm ss}}{r_{0}}\left[\frac{\ln\left[(R/r_{0})+\sqrt{1+(R/r_{0})^{2}}\right]}{(R/r_{0})}\right],
\end{equation}
with $R=\sqrt{r^{2}+z^{2}}$, radial scale size $r_{0}$, and mass $M_{\rm ss}$. The disk is modeled as a Plummer-Kuzmin potential \citep{1975PASJ...27..533M}
\begin{equation}\label{eq:phidisk}
\Phi_{\rm disk}(r,z)=-\frac{GM_{\rm disk}}{\sqrt{r^{2}+(a+\sqrt{z^{2}+b^{2}})^{2}}}
\end{equation}

The total potential is 
\begin{equation}\label{eq:totpot}
    \Phi_{\rm tot} = \Phi_{\rm disk} + \Phi_{\rm ss}.
\end{equation}

For our simulations we set the values for mass and scale lengths as shown in Table \ref{tab:parameters}. 

\begin{deluxetable}{lD}\label{tab:parameters}
\tablehead{
\colhead{Constant} & \multicolumn2c{Value}
}
\startdata
$M_{ss}$ (\msun) & 4.0x$10^9$\\
$M_{disk}$ (\msun) & 1.0x$10^{10}$\\
$e_{disk}$ & 0.95\\
$r_0$ (pc) & 350.0\\
a (pc) & 150.0\\
b (pc) & 75.0\\
$\Gamma$ & 5.0\\
$\chi$ & 5.0\\
\enddata
\caption{Values for initial conditions and the AGN jet as explained in Section \ref{sec:setup}.}
\end{deluxetable}

We model the initial gas density as a two phase medium with a hot diffuse halo and a cool disk. We use the total gravitational potential given in Equation \ref{eq:totpot} to set the density of the smooth halo using,
\begin{equation}\label{eq:halodens}
    n_{\rm halo}(r,z) =  n_{\rm halo}(0,0)\times \exp\left[-\frac{\Phi_{\rm tot}(r,z)-\Phi_{\rm tot} (0,0)}{c_{\rm s,halo}^{2}}\right].
\end{equation}

\begin{figure}
\gridline{\fig{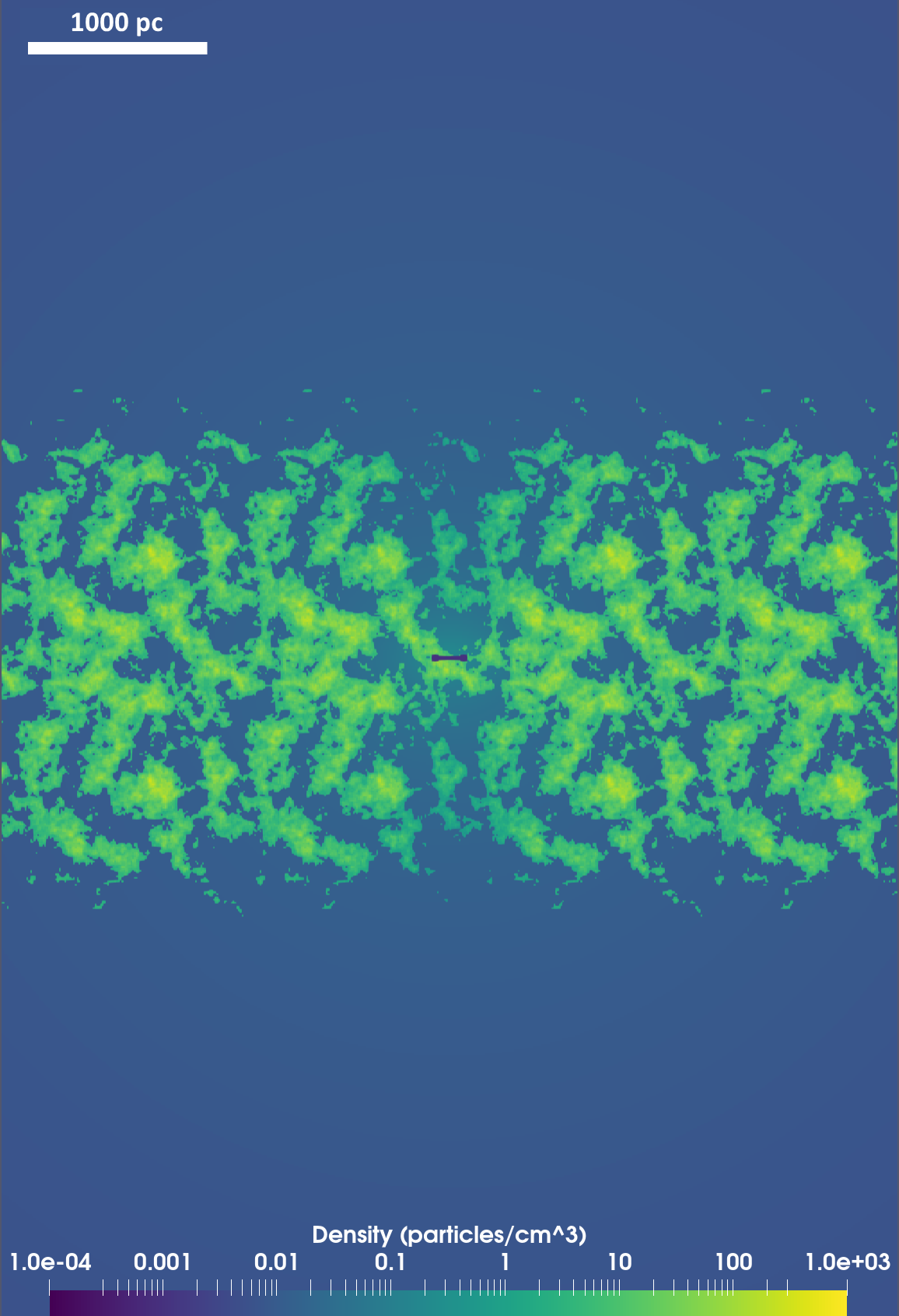}{0.33\textwidth}{}
          }
    \caption{XZ slice showing the initial density in particles cm$^{-3}$. The computational domain covers $5000\times5000\times8000$ pc divided into a grid of $500\times500\times800$ cells.}
    \label{fig:inidens}
\end{figure}

The cool disk gas is set similarly, but includes terms for rotation
\begin{equation}\label{eq:diskdens}
    n_{\rm disk}(r,z) =  n_{\rm disk}(0,0)\times \exp\left[-\frac{\Phi_{\rm tot}(r,z)-e_{\rm disk}^{2}\Phi_{\rm tot}(r,0)-(1-e_{\rm disk}^{2})\Phi_{\rm tot}(0)}{\sigma_{t}^{2}+c_{\rm s,disk}^{2}}\right]
\end{equation}
where $n(0,0)$ is the central density, $c_{s,disk} = \sqrt{k_BT_{disk}/m_H}$ the sound speed, and $e_{\rm disk}$ is the ratio of azimuthal to Keplerian velocity. 
The smooth density distribution of the disk is multiplied by a fractal distribution generated using the publicly available code pyFC written by Alex Wagner\footnote{\href{https://pypi.python.org/pypi/pyFC}{https://pypi.python.org/pypi/pyFC}}.
We apply a tanh profile to both the vertical and radial directions to constrain the disk with scale lengths of 1,000 pc and 3,000 pc respectively. 
This prevents a nonphysical disk profile at large radii as mentioned in papers that have used similar methods \citep{CooperI,Tanner17,2018MNRAS.479.5544M}. 
The disk density is then scaled to an average density of 10 $cm^{-3}$. This gives a total gas mass of $\sim 8\times10^9$ \msun and a column density of between $10^{22}$ and $10^{25}$ particles cm$^{-2}$. 
For our initial conditions, the disk gas is set to a temperature of $10^{4}$ K, and the halo gas to $5\times 10^6$ K. 
But the clumpy disk gas quickly cools to the density dependant equilibrium temperature based on the radiative cooling function used.
Figure \ref{fig:inidens} shows an XZ slice of the initial gas density.

\subsection{AGN Jet}\label{sec:setup:AGN}

At the center of the computational grid all cells within a radius of 100 pc are reset each timestep to their initial state. 
This effectively acts as an inner boundary condition on the grid. 
With our resolution the AGN region has a diameter of 20 cells, sufficient to resolve the jet cone, but not to resolve the gaseous torus around the central black hole.
Therefore we make no assumptions about the gaseous torus.
Inside the AGN region we generate a biconical outflow with an opening angle of $20^{\circ}$, based observationally on the median opening angle for a large sample of MOJAVE AGN (\cite{2017MNRAS.468.4992P}). 
The direction of the outflow can be tilted with respect to the vertical axis with $0^{\circ}$ being directed vertically out of the galaxy, and $90^{\circ}$ directed horizontally into the plane of the disk. 

Using the method employed by \cite{2011ApJ...728...29W,2016MNRAS.461..967M, 2018MNRAS.479.5544M}, which is based on \cite{1995ApJS..101...29B}, we set the pressure $(p)$ and density $(\rho)$ of the AGN jet using,

\begin{equation}\label{eq:AGNpress}
    p = P_{AGN} \frac{\gamma-1}{\gamma}\frac{1}{A\Gamma^2\frac{\sqrt{1-1/\Gamma^2}}{c}\left(1+\frac{\Gamma-1}{\Gamma}\chi\right)}
\end{equation}

and

\begin{equation}\label{eq:AGNdens}
    \rho = p\chi\frac{\gamma}{\gamma-1}.
\end{equation}

Here $P_{AGN}$ is the total AGN power, $\Gamma$ is the Lorentz factor of the jet gas, the adiabatic constant $\gamma=5/3$, $A$ is the area of the jet cone where the cone intersects the fixed sphere (radius 100 pc) at the center of the grid. 
$\chi$ is a measure of the ratio of the rest mass energy density to the enthalpy of the jet gas \citep{1995ApJS..101...29B}. 
For our simulations we set $\chi = 5.0$. This makes the assumption that the jet is mostly an electron-positron jet. 
All values for the jet and galactic gravitational potential are listed in Table \ref{tab:parameters}.

\subsection{Gas Temperature Ranges}\label{sec:setup:ranges}

For convenience in tracing the different gas components in our simulations we select a set of temperature ranges to represent the multi-phase structure of the outflows. 
The gas temperature ranges we have selected can be found in Table \ref{tab:emissionbands}. 
Some of the ranges roughly correspond some familiar emission bands. 
For example, our H$_{\alpha}$ temperature range corresponds to gas temperatures where H$_{\alpha}$ would be the strongest. 
Likewise our \hotsx, \hotmx, and \hothx temperature ranges trace different parts of the hot gas that are convenient for predicting expected X-ray emission.
This allows a rough comparison between our simulations and the expected emission in various wavebands.

In one instance (see Figure \ref{fig:00xraytime} we do calculate the total emission from the \hotsx range by calculating the total energy loss per cell based on our radiative cooling function. 
Then for Figure \ref{fig:00xraytime} we calculate the energy flux through one side of each cell and sum the fluxes along the Y axis. 
This approximates the expected soft X-ray emission, including both continuum and line emission, from our simulations, but we do not account for disk absorption or any other factors that may affect the soft X-ray emission.

\begin{deluxetable}{lD}
\tablecaption{Temperature ranges to separate different gas components.\label{tab:emissionbands}}
\tablehead{
\colhead{Name} & \multicolumn2c{Temperature Range}
}
\decimals
\startdata
Cold & $<100~$ K \\
Warm & $1,000-5,000~$ K \\
H$\alpha$ & $5,000-40,000~$ K \\
\hotsx & $0.5~-3.0~$ keV \\
\hotmx & $3.0~-10.0~$ keV \\
\hothx & $>10~$ keV
\enddata
\tablecomments{For purposes of this paper, \hotsx is hot gas in our ``soft" X-ray band; \hotmx is hot gas in our ``medium" X-ray band. \hothx is hot gas in our ``hard" X-ray band. These are for convenience in tracking different parts of the hot gas, but do not necessarily correspond to other traditional uses of soft, medium and hard X-ray bands.}
\end{deluxetable}

\section{AGN Jet Direction}\label{sec:angle}

We now consider how the direction of the AGN jet with respect to the galactic disk affects the shape, content, and velocity of the outflow. In this section we only consider simulations with $P_{AGN} = 10^{42}$ \ergs. 
At higher angles of inclination the jet will have to pass through a longer column of disk gas which does significantly affect the shape and content of the outflow. 
But we also find that the immediate structure of the ISM within 1 kpc of the AGN has as strong an effect on the outflow as the angle of jet inclination.
While it is more common for AGN jets to be roughly aligned with the minor axis of the host galaxy \citep{2005MNRAS.362...25B, 2011MNRAS.414.2148L, 2012MNRAS.421.1569B,2014ARA&A..52..589H}, there are some jets that are highly inclined with respect to the galactic axis \citep{2000ApJ...537..152K, 2003ASPC..290...47P, refId0, 2019A&A...627A..53H}. 

\begin{figure}
\gridline{\fig{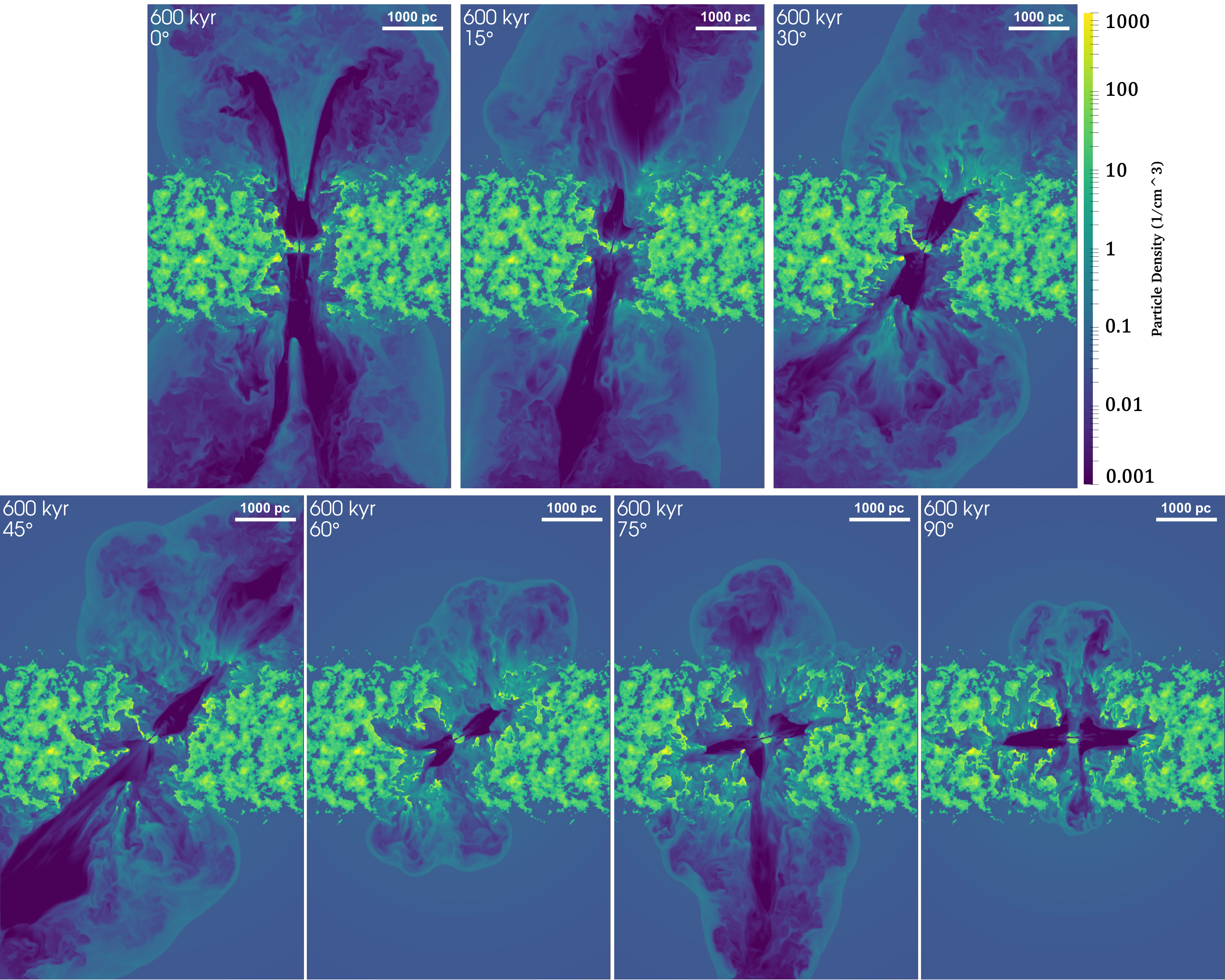}{1.0\textwidth}{}}
    \caption{XZ slice showing density in particles cm$^{-3}$. Each shows the density after 600 kyr for simulations with $P_{AGN} = 10^{42}$ \ergs. The top row shows AGN jet inclinations of $0^{\circ}$, $15^{\circ}$, and $30^{\circ}$. The bottom row shows $45^{\circ}$, $60^{\circ}$, $75^{\circ}$, and $90^{\circ}$.}
    \label{fig:600dens}
\end{figure}

In our simulations within a few hundred pc the jet forms an elongated cocoon of shocked jet gas surrounded by shocked ISM gas. 
The surrounding ISM pressure quickly collimates the jet creating a region of high velocity, laminar flow. 
This action is in agreement with more idealized jet models developed by \cite{2011ApJ...740..100B} and more recently by \cite{2019MNRAS.488.4926I}. 

At higher inclinations the jet must push through a longer ISM column requiring more energy and time before breaking out of the disk. 
As can be expected, and as shown in Figure \ref{fig:600dens}, after 600 kyr, for all angles between $0^{\circ}$ and $45^{\circ}$ either the jet or the mixed ISM jet gas has has cleared the disk and reached computational domain. 
For higher angles the jet has not left the disk, though the expanding shocked ISM gas has escaped the disk.
At $90^{\circ}$, with the jet directly into the disk, we did not expect the jet to leave the disk, but hot buoyant bubbles of shocked ISM gas have managed to expand outside of the disk. 
\cite{2019A&A...627A..53H} observed such a galactic outflow, with the AGN jet directed nearly parallel with the disk. 
The AGN dominated the central $\sim 1$ kpc and drove a multi-phase outflow similar to a starburst driven outflow \citep{Tanner17}.

\begin{figure}
\gridline{\fig{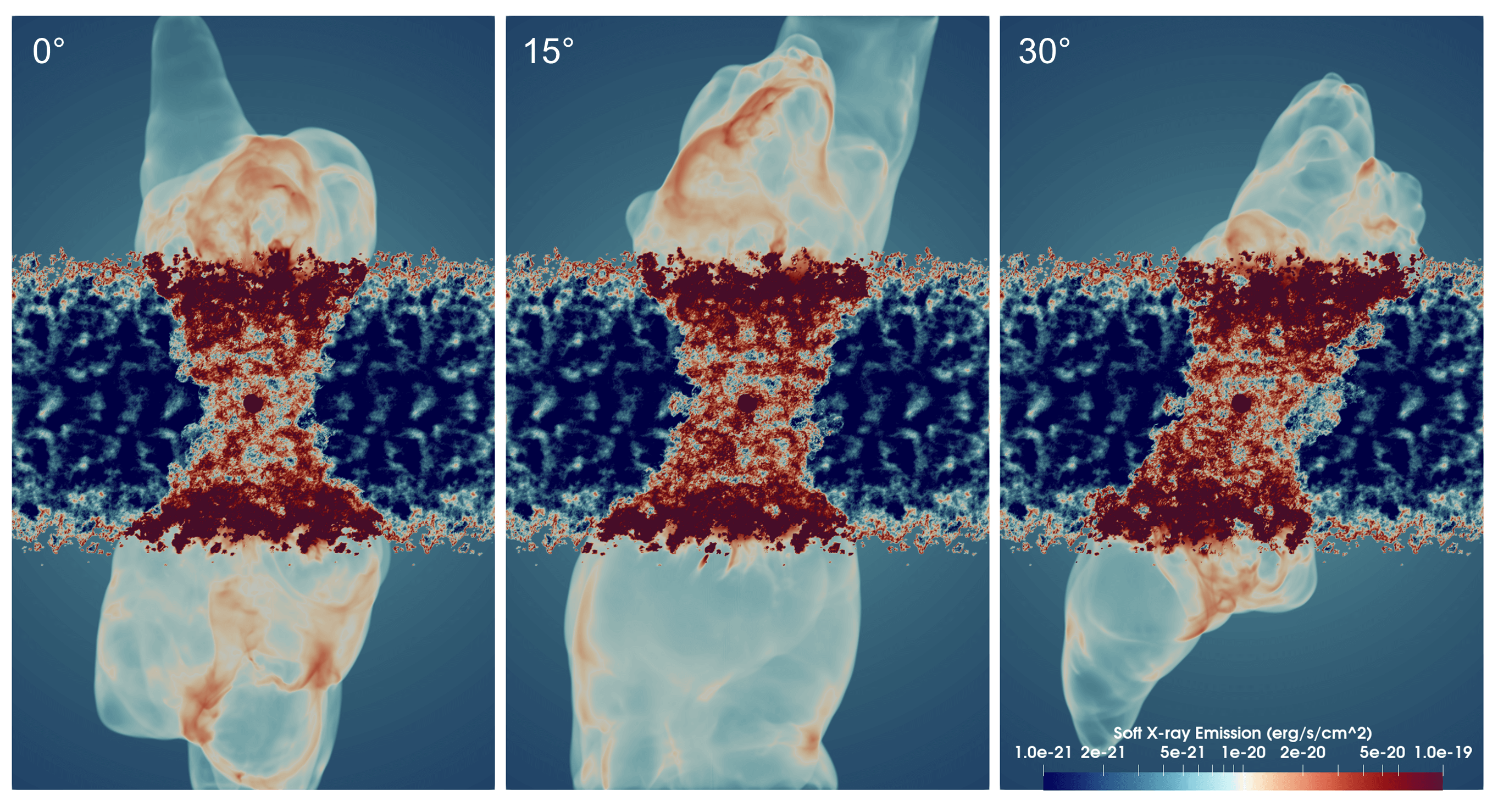}{1.0\textwidth}{}
          }
\gridline{
          \fig{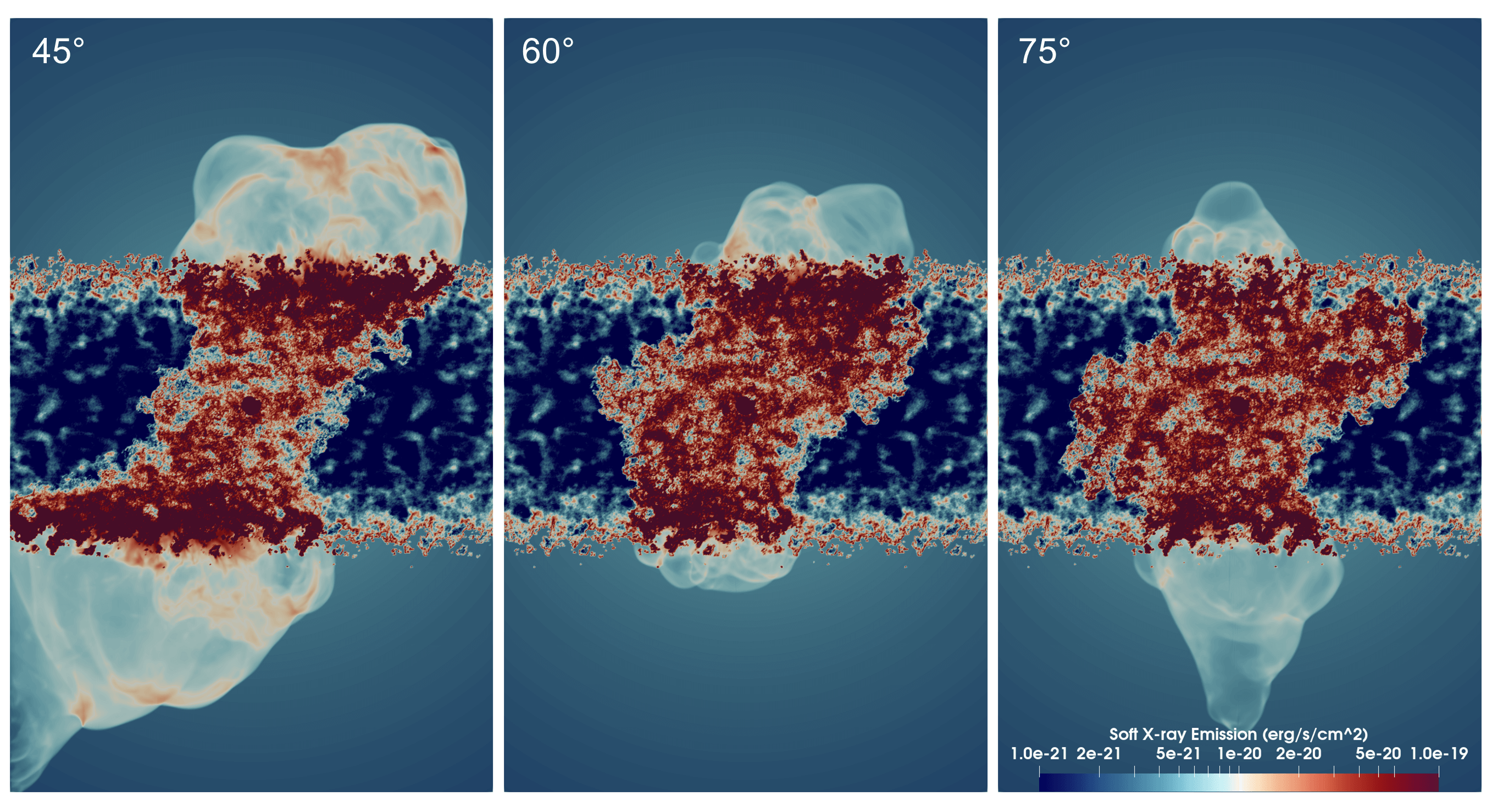}{1.0\textwidth}{}
          }
    \caption{Simulated soft x-ray emission for AGN inclination angles $0^{\circ}$ through $75^{\circ}$, with $P_{AGN} = 10^{42}$ \ergs, at 300 kyr. A shock front and a cocoon travel with the jet, but where the jet impacts a dense cloud a hot bubble forms and expands out of the disk of the galaxy. The soft X-ray emission is calculated using the radiative cooling function, then divided by six to get the flux through one cell face, and then summed along the y axis. We do not account for absorption by the disk where most of the X-rays would be absorbed.}
    \label{fig:00xraytime}
\end{figure}

While the inclination of the jet does dominate the evolution of low power jet, with more highly inclined jet angles creating smaller outflows, there are a few interesting variations as part of this general trend. 
For example, with the jet at $75^{\circ}$ part of the jet is deflected downward $\approx 80^{\circ}$ from the original jet direction by a large, dense cloud (Figure \ref{fig:600dens}; $75^{\circ}$). 
This makes the outflow extremely asymmetric. 
To approximate the soft X-ray emission we calculate the energy flux from each cell using the radiative cooling function for cells with a temperature in the \hotsx range. 
This helps us trace the expected soft X-ray emission from the shells of hot gas surrounding the jets and from shock heated ISM gas in the disk. 
While there is X-ray emission on both sides of the galaxy (Figure \ref{fig:00xraytime}; $75^{\circ}$), only the bottom side would have gas traveling with high enough velocity to generate significant synchrotron radio emission. 
Due to the deflected jet, a larger X-ray bubble is formed than compared to those formed with the jet at $60^{\circ}$ (see Figure \ref{fig:00xraytime}; $60^{\circ}$ and $75^{\circ}$).

At $30^{\circ}$ the jet in both directions encounters dense clouds and does not have a clear path out of the disk (Figure \ref{fig:600dens}; $30^{\circ}$). 
This causes the jet on the bottom side of the galaxy to deflect slightly, but on the top side the jet barely penetrates the edge of the disk. 
This contrasts with the jet at $45^{\circ}$ which manages to bypass the same clouds that stopped the jet at $30^{\circ}$. 
While a jet angle at $30^{\circ}$ has a shorter path out of the disk, the jet at $45^{\circ}$ forms larger X-ray bubbles, as seen in Figure \ref{fig:00xraytime}.

\begin{figure}
\gridline{
          \fig{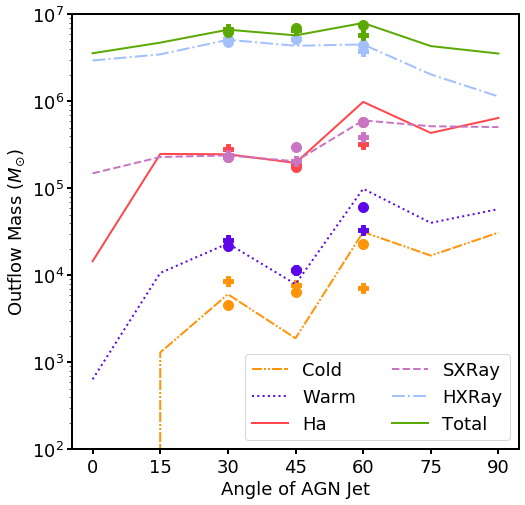}{0.5\textwidth}{}
          \fig{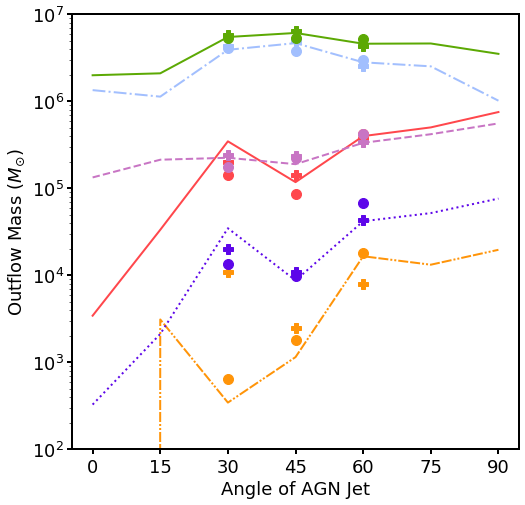}{0.5\textwidth}{}
          }
    \caption{Mass above the escape velocity in different temperature bands. Temperature bands can be found in Table \ref{tab:emissionbands}. The left graph shows total outflow mass in the -z direction at 600 kyr, and the right graph shows in the +z direction. In these graphs we omit data for the \hotmx mass for clarity since it is very similar to both the \hotsx and H $\alpha$ masses.}
    \label{fig:outmassangle}
\end{figure}

\begin{figure}
\gridline{\fig{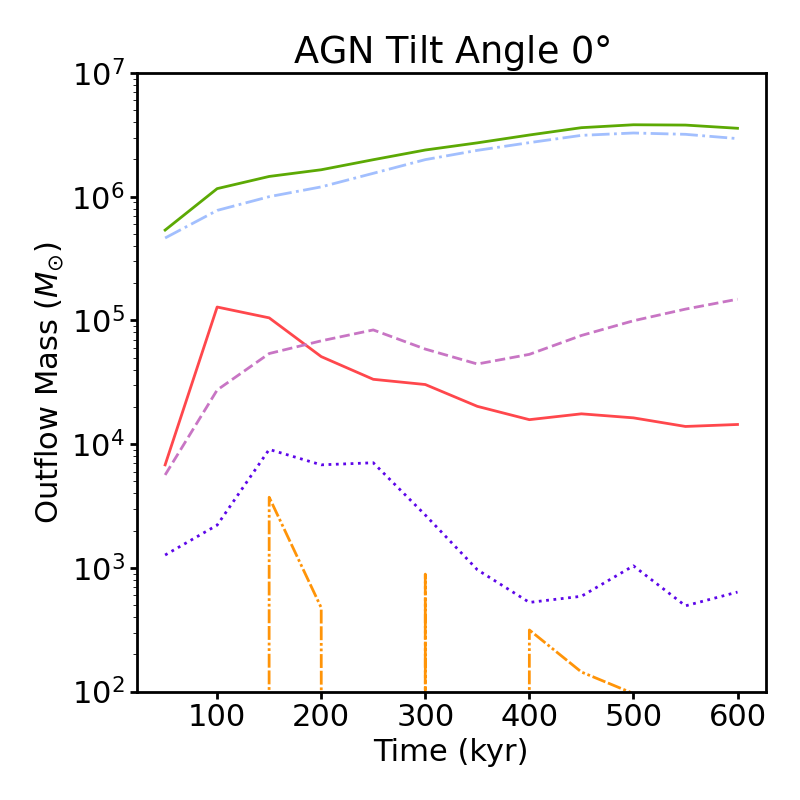}{0.33\textwidth}{}
          \fig{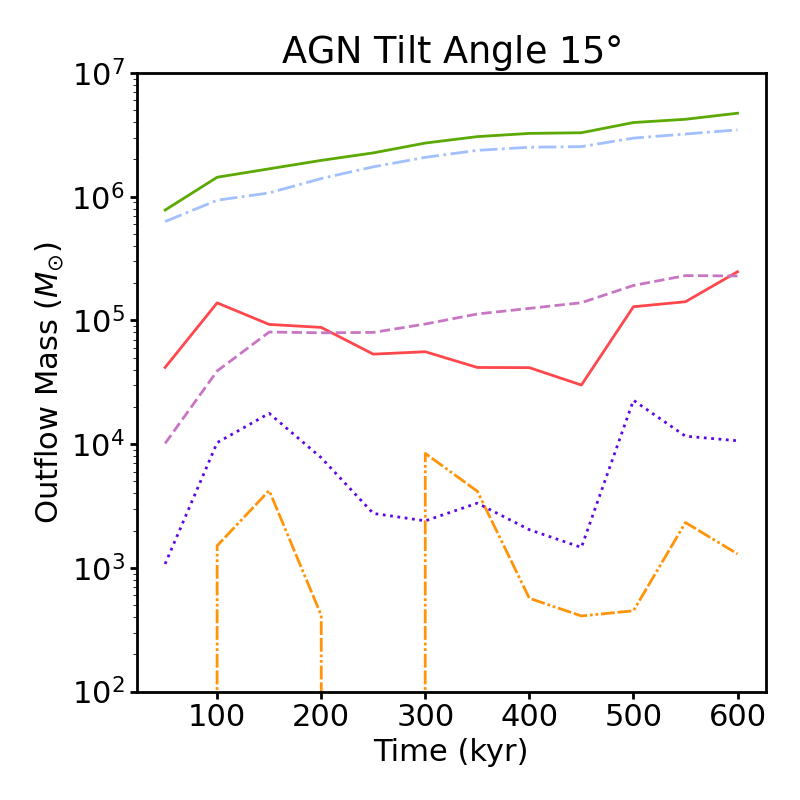}{0.33\textwidth}{}
          \fig{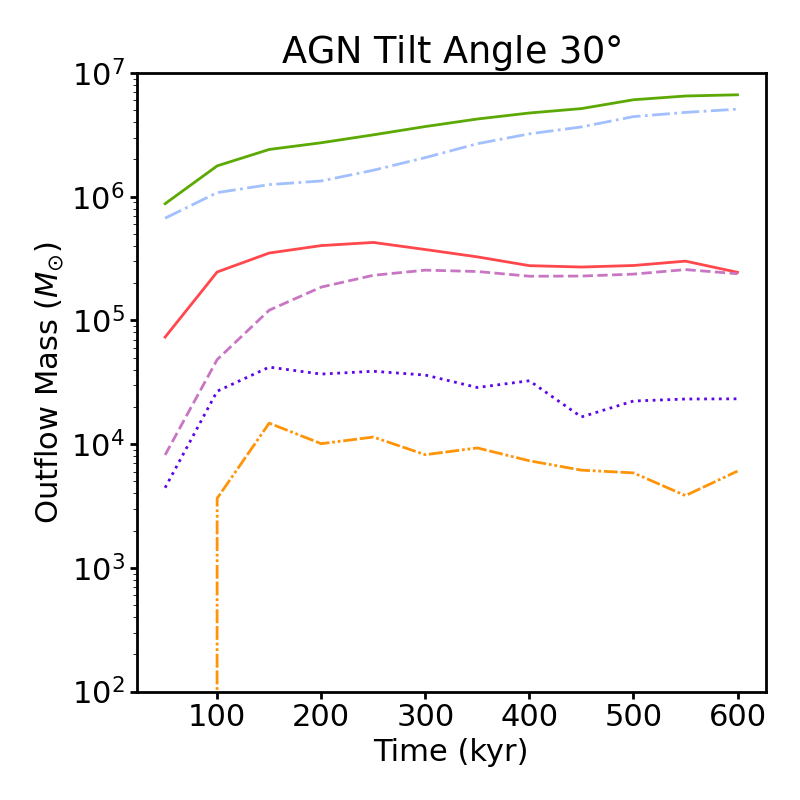}{0.33\textwidth}{}
          }
\gridline{
          \fig{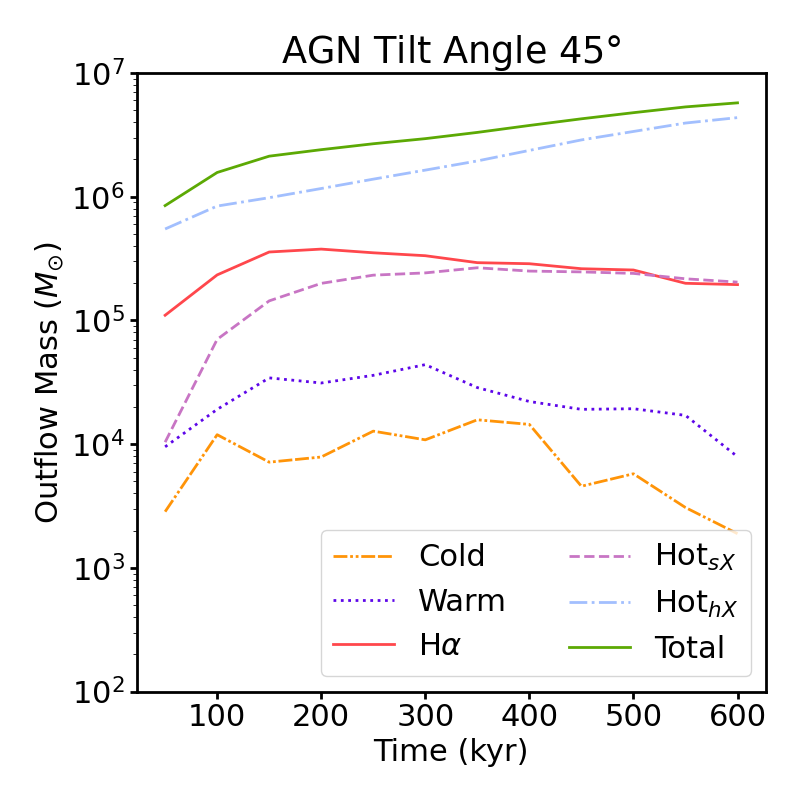}{0.33\textwidth}{}
          \fig{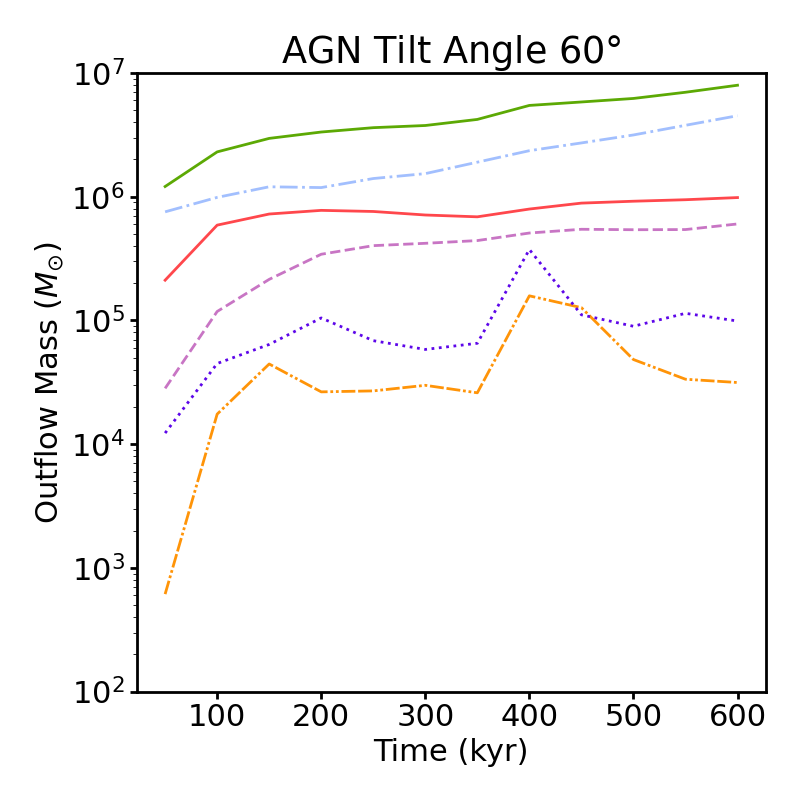}{0.33\textwidth}{}
          \fig{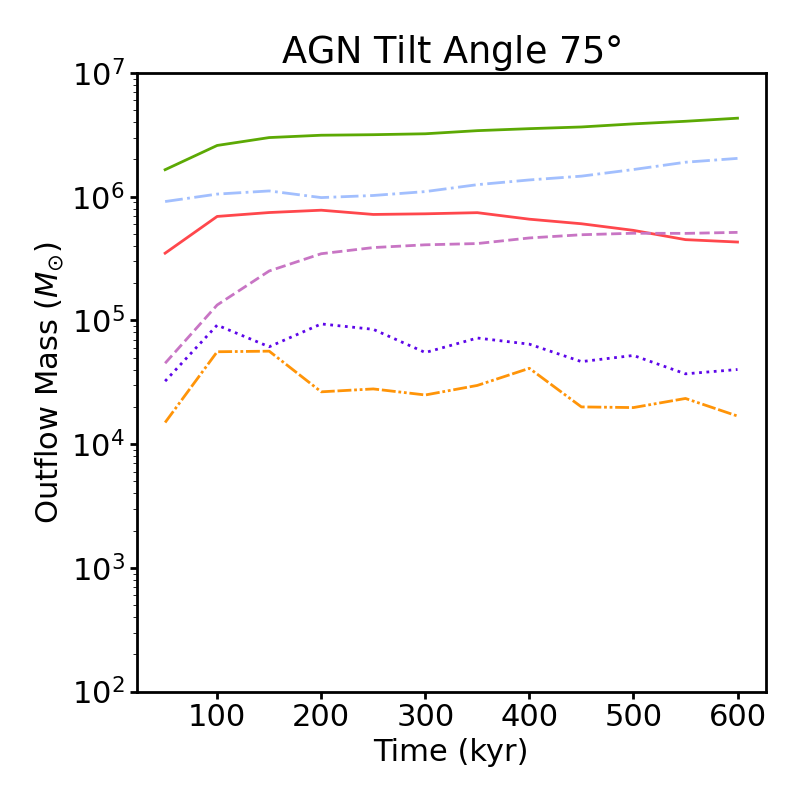}{0.33\textwidth}{}
          }
\gridline{
          \fig{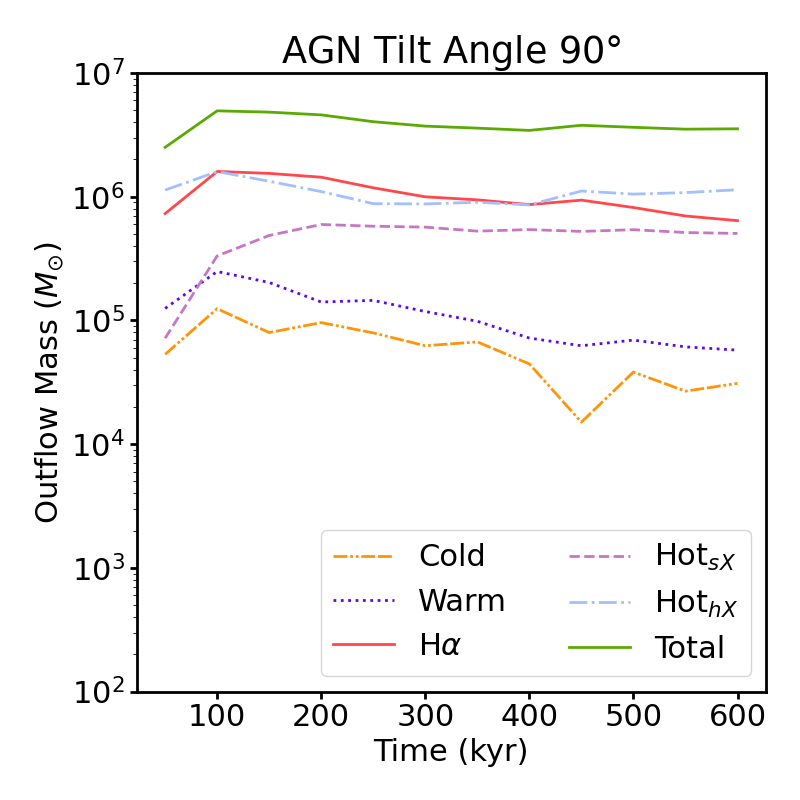}{0.33\textwidth}{}
          }
    \caption{Mass above the escape velocity in different temperature bands. Temperature bands can be found in Table \ref{tab:emissionbands}. Lines show escape mass for simulations with a luminosity of $10^{42}$ erg s$^{-1}$. Simulations with luminosities of $10^{41}$ erg s$^{-1}$ and $10^{43}$ erg s$^{-1}$ are shown with circles and crosses respectively. Markers are colored according to the emission band. The other graphs show how the mass above the escape velocity changes during the simulation time. In some of the simulations at certain timesteps there was no cold gas above the escape velocity. In the graphs above this is the reason for the vertical lines in the cold gas.}
    \label{fig:outmass}
\end{figure}

To understand how the velocity differs for the temperature ranges we calculate an escape velocity for each cell in the simulation using the gravitational potential (Equation \ref{eq:totpot}). 
We then bin the cells according to the their velocity and sum up the mass in each velocity bin, separating the mass by temperature using the temperature ranges given in Table \ref{tab:emissionbands}. 
Then we sum the mass for all cells with a velocity above the escape velocity for that cell. 
In Figure \ref{fig:outmassangle} we show the total mass in each temperature range at 600 kyr with respect to the AGN tilt angle. 
Then in Figure \ref{fig:outmass} we show how the out flowing mas above the escape velocity changes over the course of the simulation. 

As shown in Figure \ref{fig:outmassangle} for all AGN angles the total outflow mass at 600 kyr is of the same order of magnitude between $10^6-10^7$ \msun, with the highest values at angles $30^{\circ}$ - $60^{\circ}$. 
In our simulations the total initial gas mass is $\approx 9\times 10^9$ \msun. 
Thus at 600 kyr between $0.1 \%$ and $1.0 \%$ of the total initial gas mass has a velocity above the escape velocity. 
Whether this gas is completely removed from the galaxy and galactic halo will depend on factors such as the circumgalactic environment outside of our simulations. 

\begin{figure}
\gridline{\fig{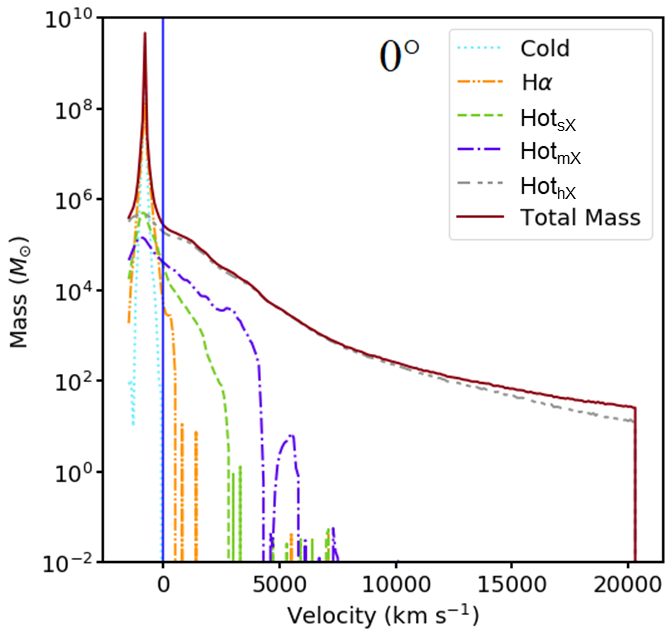}{0.33\textwidth}{}
          \fig{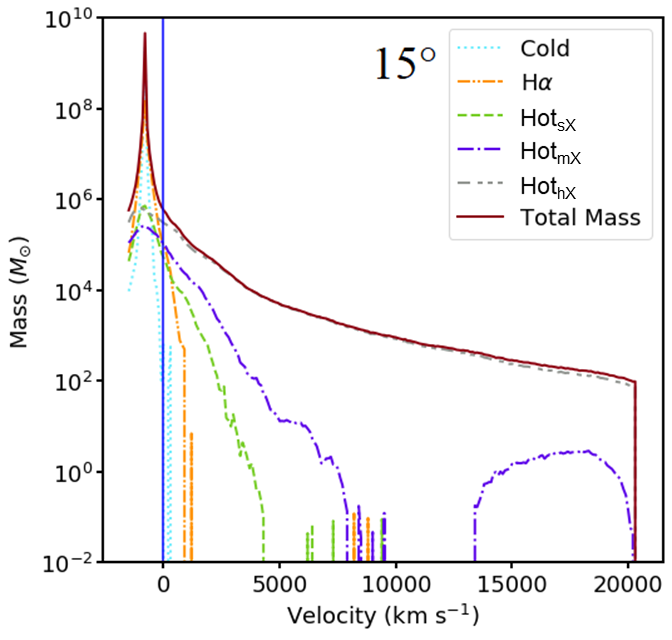}{0.33\textwidth}{}
          \fig{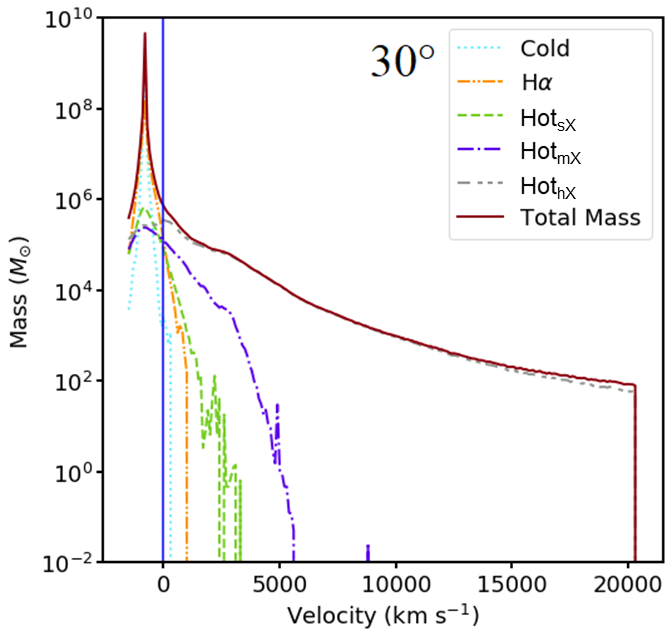}{0.33\textwidth}{}
          }
\gridline{
          \fig{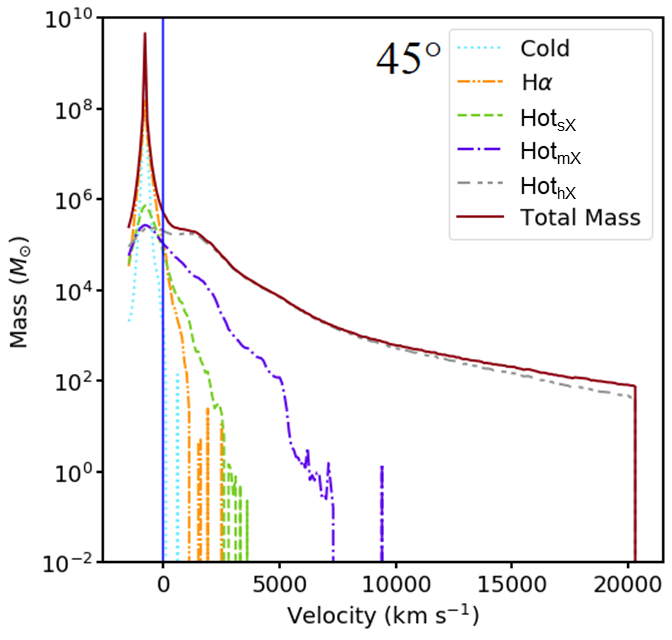}{0.33\textwidth}{}
          \fig{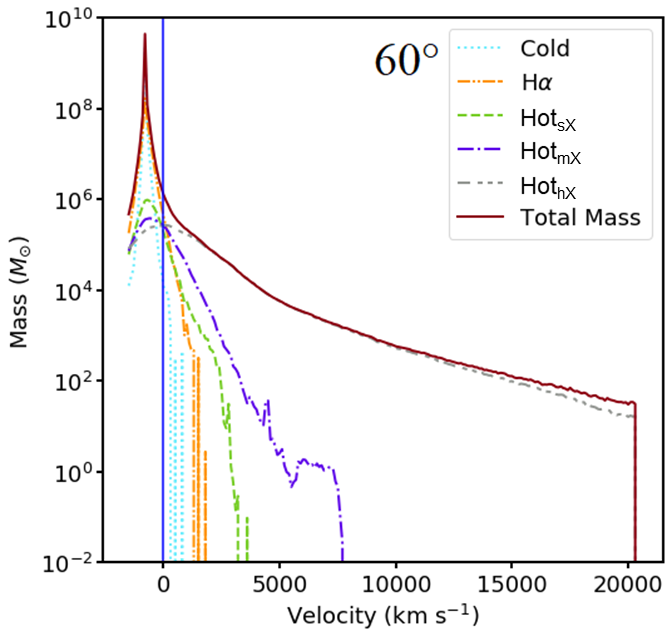}{0.33\textwidth}{}
          \fig{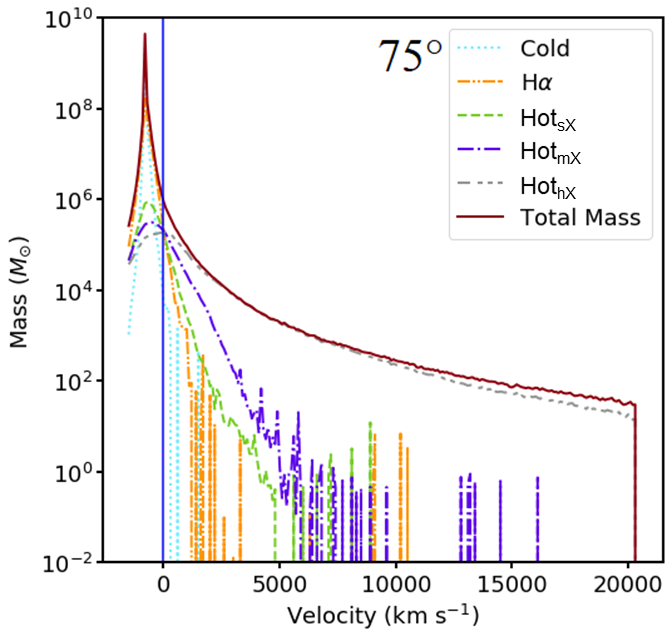}{0.33\textwidth}{}
          }
\gridline{
          \fig{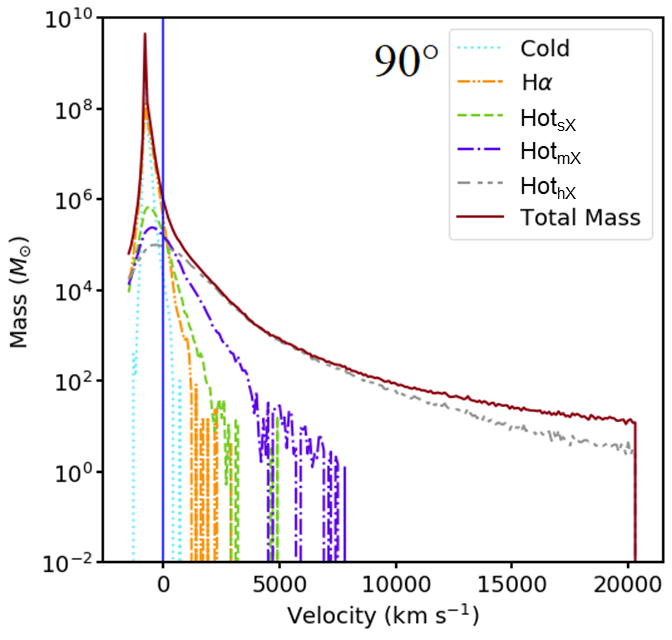}{0.33\textwidth}{}
          }
    \caption{Probability distribution functions (pdf) of the mass with respect to the escape velocity in simulations with $P_{AGN} = 10^{42}$ \ergs. The vertical line at 0 \kms is the escape velocity of the gas. To the right of the line is gas above the escape velocity, to the left below the escape velocity. The peak in each graph corresponds to the rest frame of the galaxy. This shows the state of the gas at 600 kyrs. Binning is cut off at 20,000 \kms.}
    \label{fig:masspdf}
\end{figure}

As can be seen Figure \ref{fig:outmassangle}, for simulations with shallow angles of inclination $> 95\%$ of the total mass of the outflow is \hothx ($> 10^8$ K). 
While for higher angles of inclination the \hothx band makes up a smaller fraction of the gas above the escape velocity. 
At $90^{\circ}$ the mass of the \hothx gas is of the same order of magnitude with H$\alpha$ emitting gas. 
Also at all angles the cold and warm gas makes up $\lesssim 1\%$ of the total outflow mass, but with an increasing AGN angle the total amount of cold and warm gas mass increases by two orders of magnitude. 
In Figure \ref{fig:outmass} we show that the cold, warm, and H$\alpha$ emitting gas tends to decrease over time as it is shock heated, with a corresponding increase in hot gas (both \hotsx and \hothx).

In Figure \ref{fig:masspdf} we plot the probability distribution functions (pdf) of mass versus outflow velocity for all angles at 600 kyrs. 
In the graphs we set $v = 0$ \kms (indicated by a vertical line) to the escape velocity calculated for each cell. 
To the right of the vertical line is mass above the escape velocity, to the left below the escape velocity. 
In all cases the \hothx gas dominates the high speed outflows. 
As shown in Figure \ref{fig:outmassangle}, at higher angles the total outflow mass for both the H$\alpha$ emitting and \hotsx gas may be approximately the same order of magnitude, but as seen from \ref{fig:masspdf} the \hotsx gas consistently has a much higher velocity.

We do see the same behavior as shown in Figure \ref{fig:outmass} with the outflow becoming more multi-phase at higher angles of AGN inclination. 
By looking just to the left of the vertical lines in Figure \ref{fig:masspdf} we see a significant amount of cold and warm gas just below the escape velocity. 
This gas consists of dense clouds still moving upwards but will most likely dissipate before reaching the CGM. 
If we just consider the gas with velocities around the escape velocity then for all angles $m_{Hot_{sX}} \approx m_{cold}$ and $m_{Hot_{hX}} \approx m_{H\alpha}$.

As shown in Figures \ref{fig:outmass} and \ref{fig:masspdf} the angle of inclination does not significantly affect the total mass in the outflow, but it does significantly affect the composition of the outflow. 
Higher angles of inclination have a more multi-phase outflow. 
This is because at lower angles the outflow is dominated by the hot, momentum-driven jet, while at higher angles it is dominated by the pressure-driven, shocked ISM gas mixed with jet gas. 
Thus the pressure-driven bubbles are much more efficient at lifting cold gas off of the disk than the momentum-driven jets.

\section{AGN Power}\label{sec:AGNpower}

It is logical to assume that the AGN power should determine the morphology of the resulting outflow. 
More powerful AGNs do produce more extensive jets \citep[for example,][]{2021A&ARv..29....3O}. 
But based on our results in Section \ref{sec:angle} we found that for low power AGNs the ISM can significantly affect the morphology of the outflow. 
To determine how the AGN power affects the morphology of the galactic outflow we now look at simulations with $P_{AGN}$ ranging over six orders of magnitude. 
This covers a range of AGNs from Seyfert galaxies to quasar galaxies. 

\begin{figure}
\gridline{
          \fig{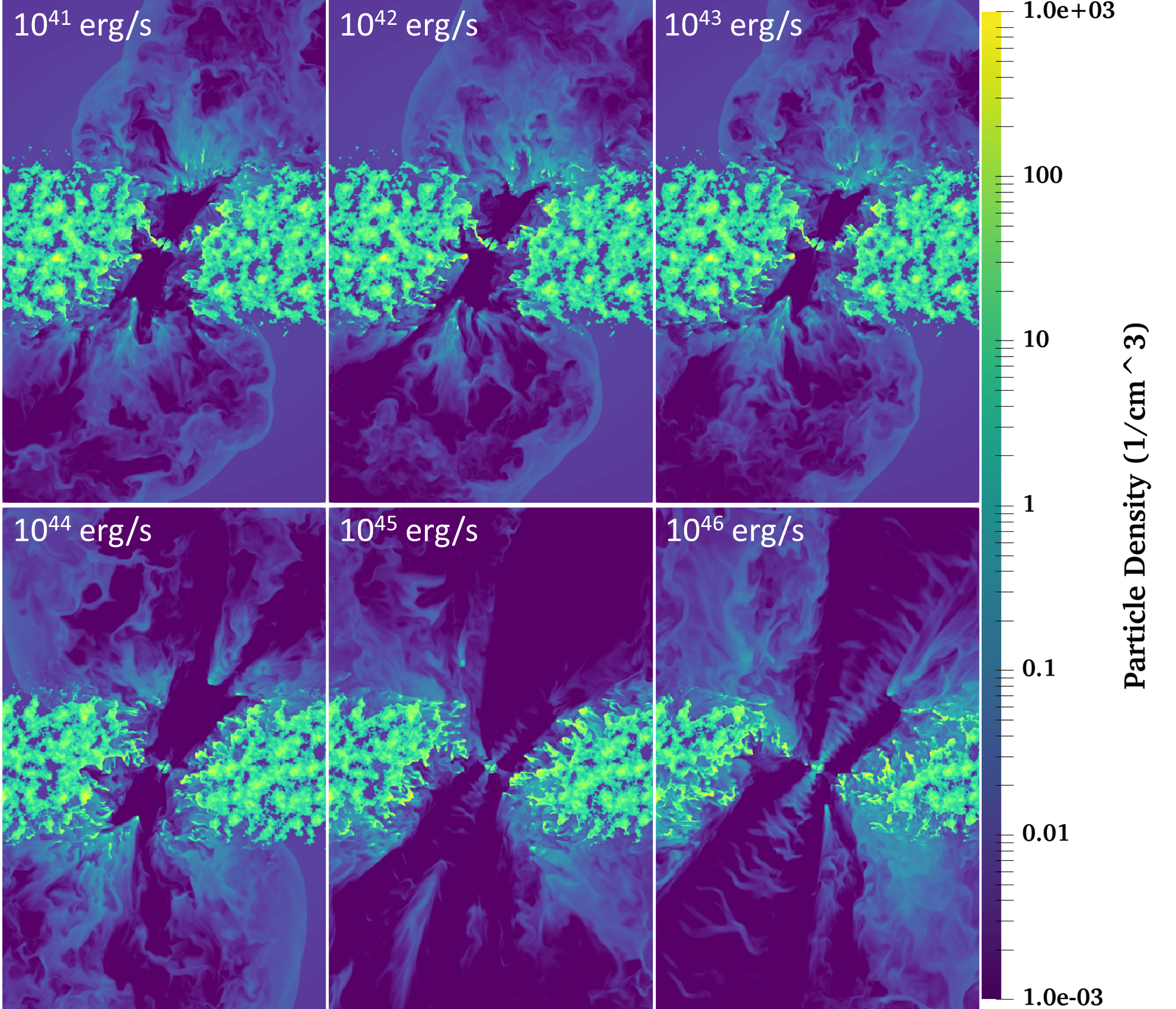}{1.0\textwidth}{}
         }
\caption{XZ slices showing the density for simulations with $30^{\circ}$ for $P_{AGN} = 10^{41}-10^{46}$ \ergs.\label{fig:30power}}
\end{figure}

In Figure \ref{fig:30power} we show XZ slices of the density for simulations with an AGN inclination of $30^{\circ}$ and powers ranging from $10^{41}$ \ergs to $10^{46}$ \ergs. 
For $P_{AGN} <= 10^{44}$ \ergs the ISM acts as a significant barrier to creating a free-flowing outflow. 
Even after 600 kyrs, there are still dense clouds in the path of the jet significantly disrupting the outflow. 
But for $P_{AGN} >= 10^{45}$ \ergs these clouds have been cleared from the path of the jet and have almost entirely dissipated into the hot bubble of the outflow.

\begin{figure}
\gridline{
          \fig{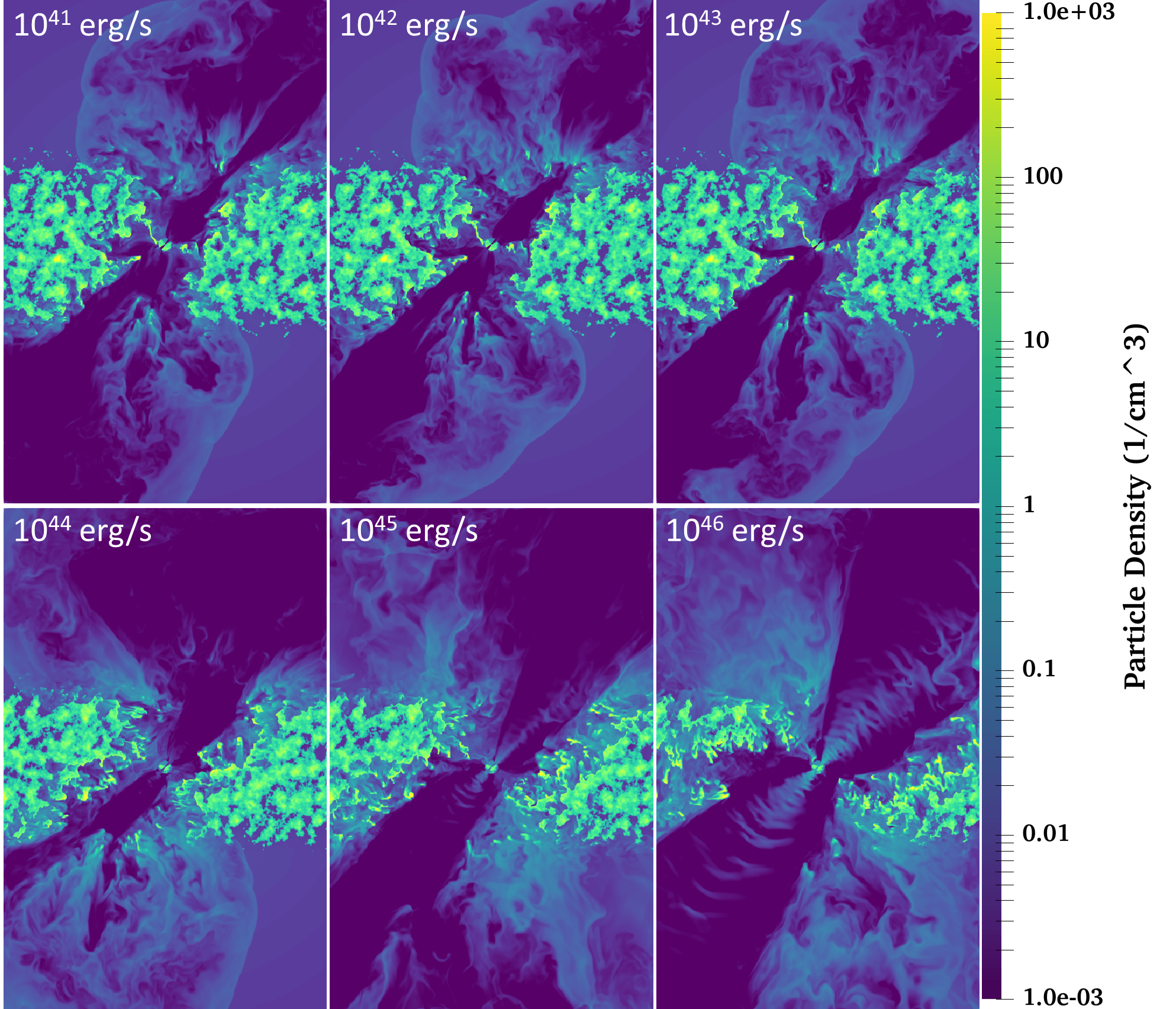}{1.0\textwidth}{}
         }
\caption{XZ slices showing the density for simulations with $45^{\circ}$ for $P_{AGN} = 10^{41}-10^{46}$ \ergs.\label{fig:45power}}
\end{figure}

In Figure \ref{fig:45power} we show the same thing as in Figure \ref{fig:30power} except with an AGN inclination angle of $45^{\circ}$. 
As before for $P_{AGN} <= 10^{44}$ \ergs the ISM acts as a significant barrier to creating a free-flowing outflow. 
For the simulation with $P_{AGN} = 10^{44}$ \ergs the jet has managed to clear a path out of the top of the galaxy but not through the bottom. 

To understand why this is we will consider a simple order of magnitude calculation to derive the energy input needed for a jet to clear a pathway through the ISM. 
Assuming a coupling efficiency of $\sim$ 5\% \cite{} over the 600 kyrs of the simulation an AGN with $P_{AGN} = 10^{44}$ \ergs should contribute $\sim 10^{56}$ ergs to the ISM. 
If we consider the column of ISM gas within 300 pc of the center line of the jet through the disk, the total mass in this column has a maximum value of $\sim 10^7$ \msun with a minimum of $\sim 10^4$ \msun depending on the distribution of dense clouds in our simulations. 
The energy required to lift $\sim 10^7$ \msun off of the disk and out of the gravitational potential of the host galaxy is $\sim 10^{56}$ ergs, for the galaxy mass as given in Table \ref{tab:parameters}. 
With the exception AGN inclination angles $\sim 90^{\circ}$, $\sim 10^{56}$ ergs can be considered a practical upper limit on the total energy needed by an AGN to clear a path through the ISM.

This means that for almost all AGNs with $P_{AGN} > 10^{44}$ \ergs the jet will impart enough energy to the ISM to remove all the gas along the path of the jet regardless of the ISM structure. 
For AGNs with $P_{AGN} < 10^{44}$ \ergs whether or not the jet can clear a path will depend on the structure of the ISM, the total gas mass along the path of the jet, and the temporal duration of the jet. 
For a lower power AGN to clear a path through the ISM it must last for longer than the 600 kyr considered in our simulations.
Such a case has been considered by \citet{2021ApJ...922..254C} in simulating the Fermi bubbles in the Milky Way and in NGC 1068.
In their simulations they assumed an intermittent AGN with $P_{AGN} = 10^{41}$ \ergs lasting for 8 Myr. 
With a total energy input of $\sim 10^{55}$ ergs the AGN jet considered by \citet{2021ApJ...922..254C} would be sufficient to clear a path through all but the densest ISMs.

\section{Jets and Bubbles}\label{sec:jetbub}

On the simplest level the outflows have two components. 
\begin{enumerate}
\item A high velocity ($> 0.1$ c) momentum-driven jet of gas with smooth, laminar flow.
\item A lower velocity, turbulent bubble of pressure supported expanding hot gas consisting of mixed ISM and jet gas.
\end{enumerate}

The growth of the high velocity jet depends on its interaction with the ISM. 
If the jet has a relatively free path through low density voids in the ISM then it forms a well collimated laminar outflow, surrounded by a turbulent cocoon. If a sufficiently large cloud is in the path of the jet then the jet can deflect, split, or stop it altogether.
For the pressure bubble, any interaction with the dense ISM drives the growth of the hot, turbulent, bubble surrounding the jet. 
The dense ISM significantly decreases the velocity of the jet gas and causes instabilities to grow in the turbulent mixing region surrounding the high velocity gas.

\begin{figure}
\gridline{\fig{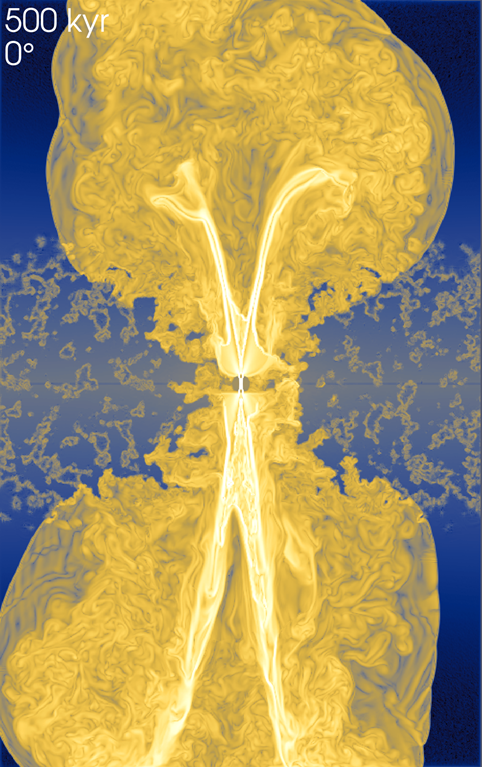}{0.25\textwidth}{}
          \fig{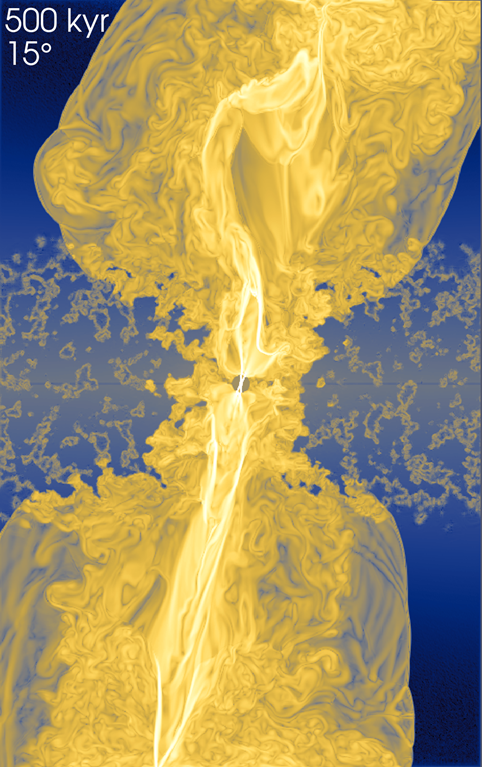}{0.25\textwidth}{}
          \fig{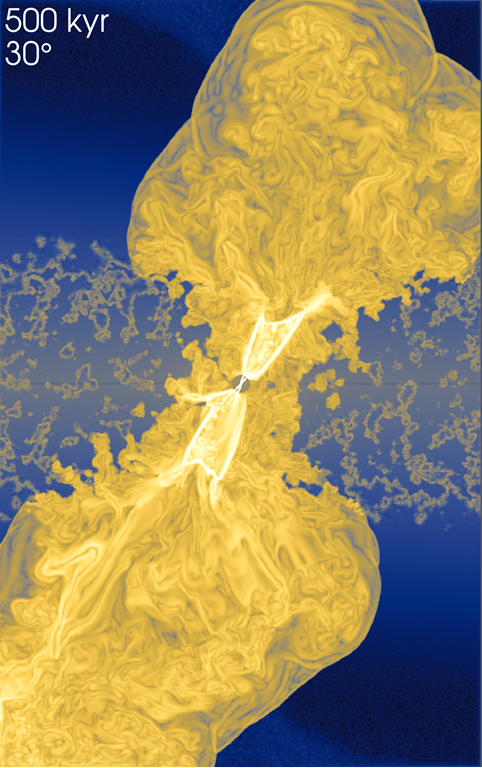}{0.25\textwidth}{}
          }
\gridline{
          \fig{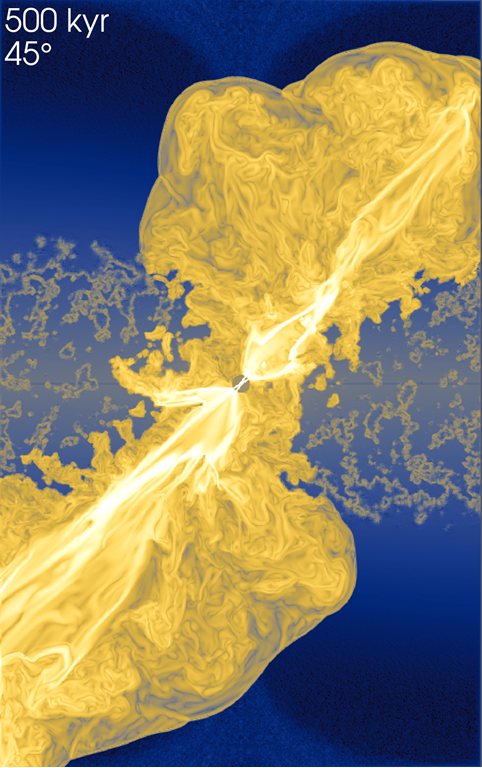}{0.25\textwidth}{}
          \fig{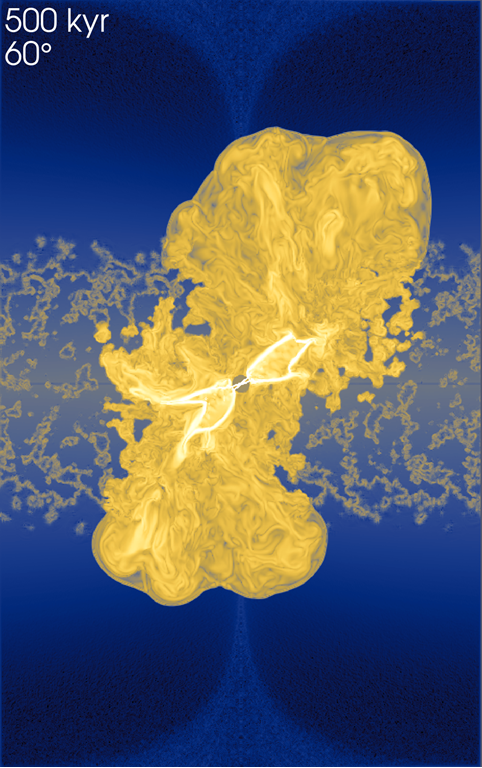}{0.25\textwidth}{}
          \fig{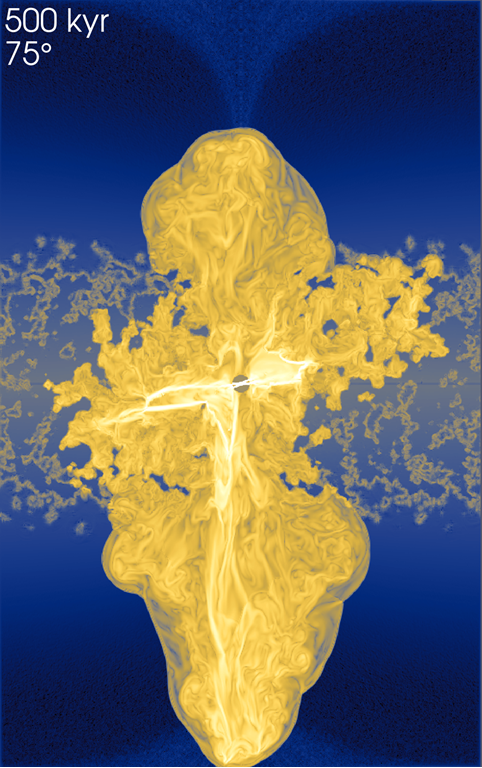}{0.25\textwidth}{}
          \fig{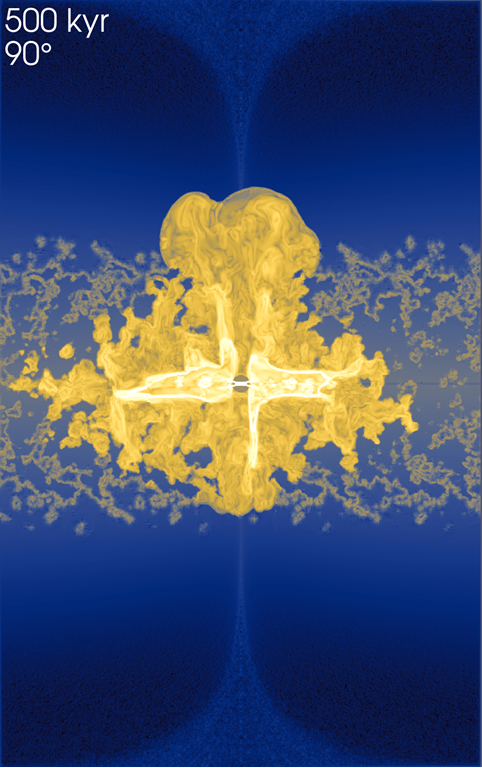}{0.25\textwidth}{}
          }
    \caption{The magnitude of the vorticity for simulations with $P=10^{42}$ \ergs in the XZ plane at 500 kyr. Each simulation has a turbulent cloud that escapes the disk, but not all have a laminar jet escaping the disk. With a jet angle at $30^{\circ}$ the jet is blocked on both sides by dense clouds, but at $45^{\circ}$ the jets miss the clouds and escape the disk. At $75^{\circ}$ a jet is deflected downwards.}
    \label{fig:vorticity}
\end{figure}

In Figure \ref{fig:vorticity} we show the magnitude of the vorticity, or curl ($\mathbf{\omega} = \nabla \times \mathbf{v}$), of the gas in the XZ plane at 500 kyr. 
The boundaries of the high velocity jets can be clearly seen due to the large velocity difference with the surrounding gas, resulting in high values for the magnitude of the curl.
Surrounding the high velocity jets is the turbulent bubble, with a thin shell ($< 100$ pc) at the shock front of the expanding bubble.

With the AGN angle at $30^{\circ}$ the laminar part of the outflow fails to leave the galactic disk because of interactions with the dense ISM. 
With a slight shift to $45^{\circ}$ the jet bypasses the clouds that disrupted the jet at $30^{\circ}$. 
Because the jet at $45^{\circ}$ still interacts with the same clouds, the hot turbulent bubbles for both $30^{\circ}$ and $45^{\circ}$ are still approximately the same size after 500 kyr. 

\begin{figure}
\gridline{\fig{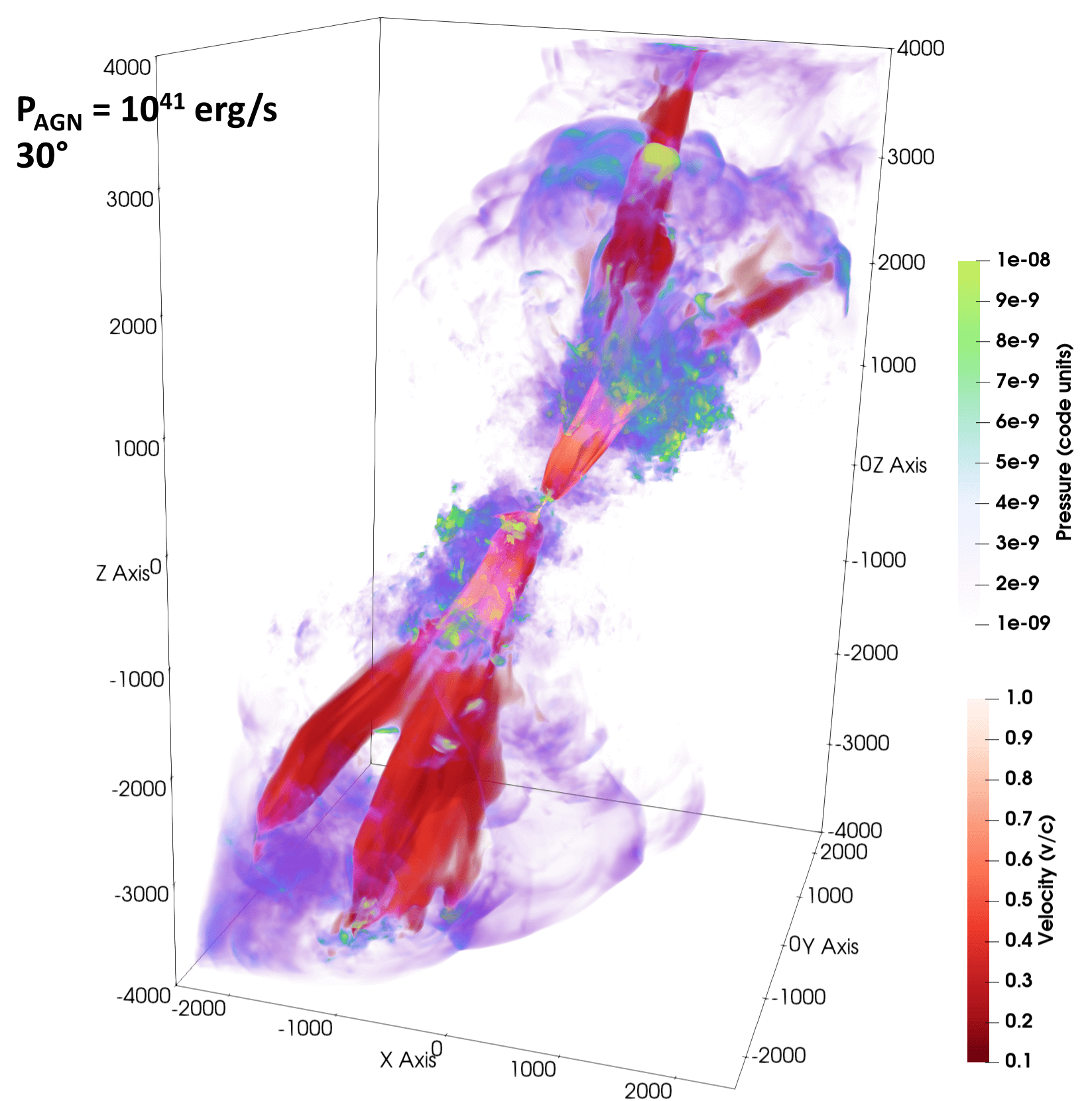}{0.33\textwidth}{}
          \fig{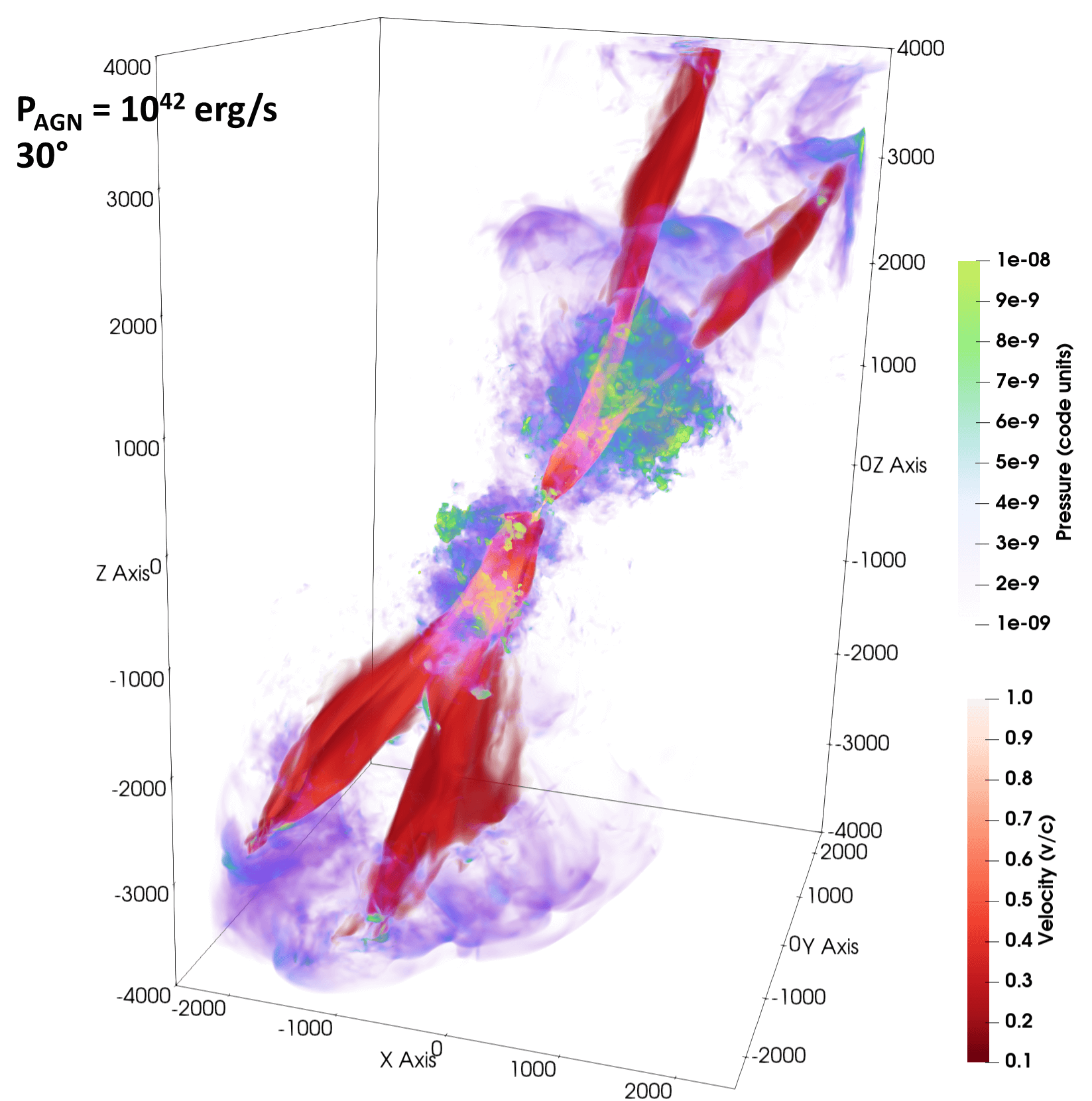}{0.33\textwidth}{}
          \fig{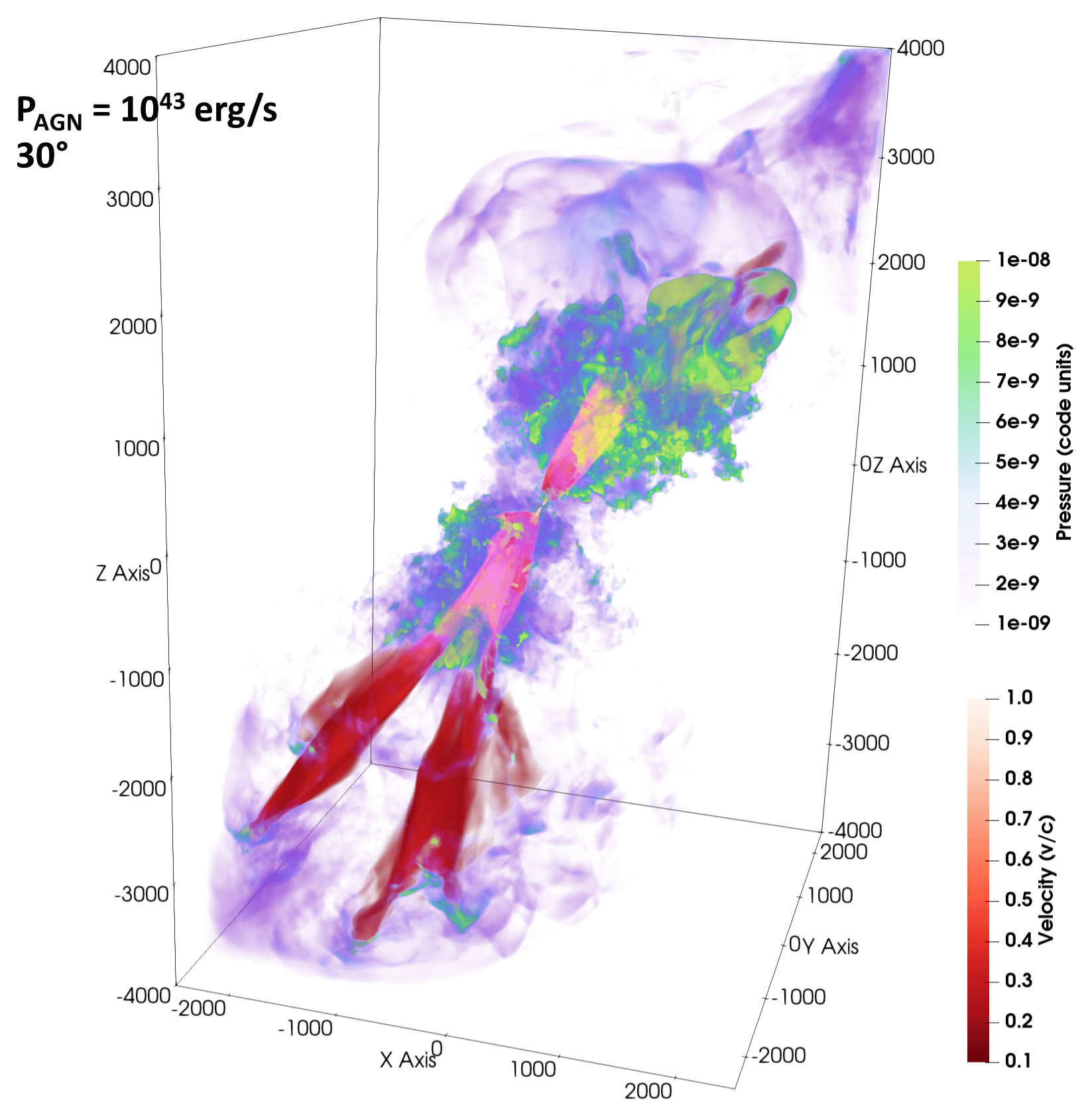}{0.33\textwidth}{}
          }
\gridline{
          \fig{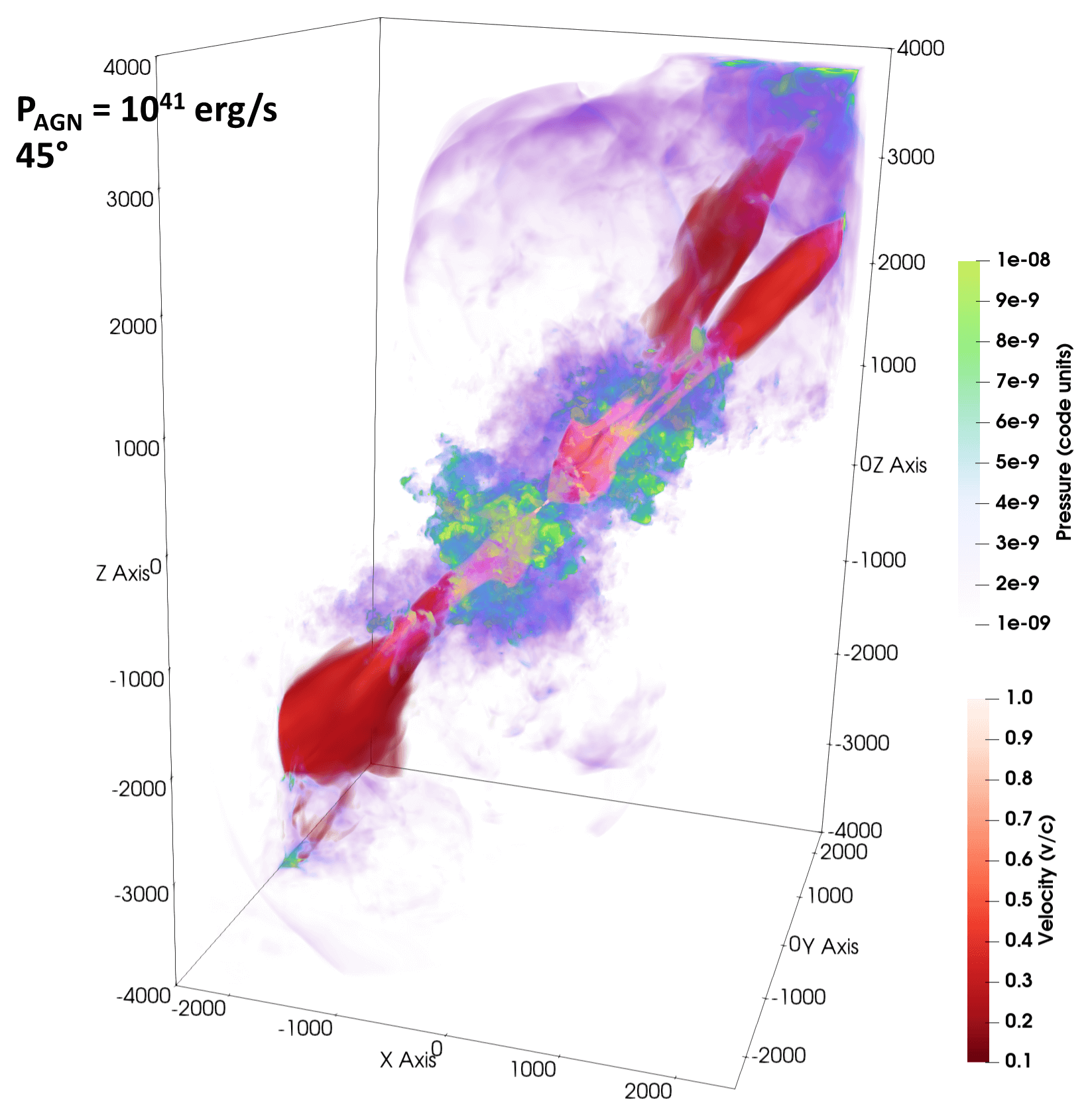}{0.33\textwidth}{}
          \fig{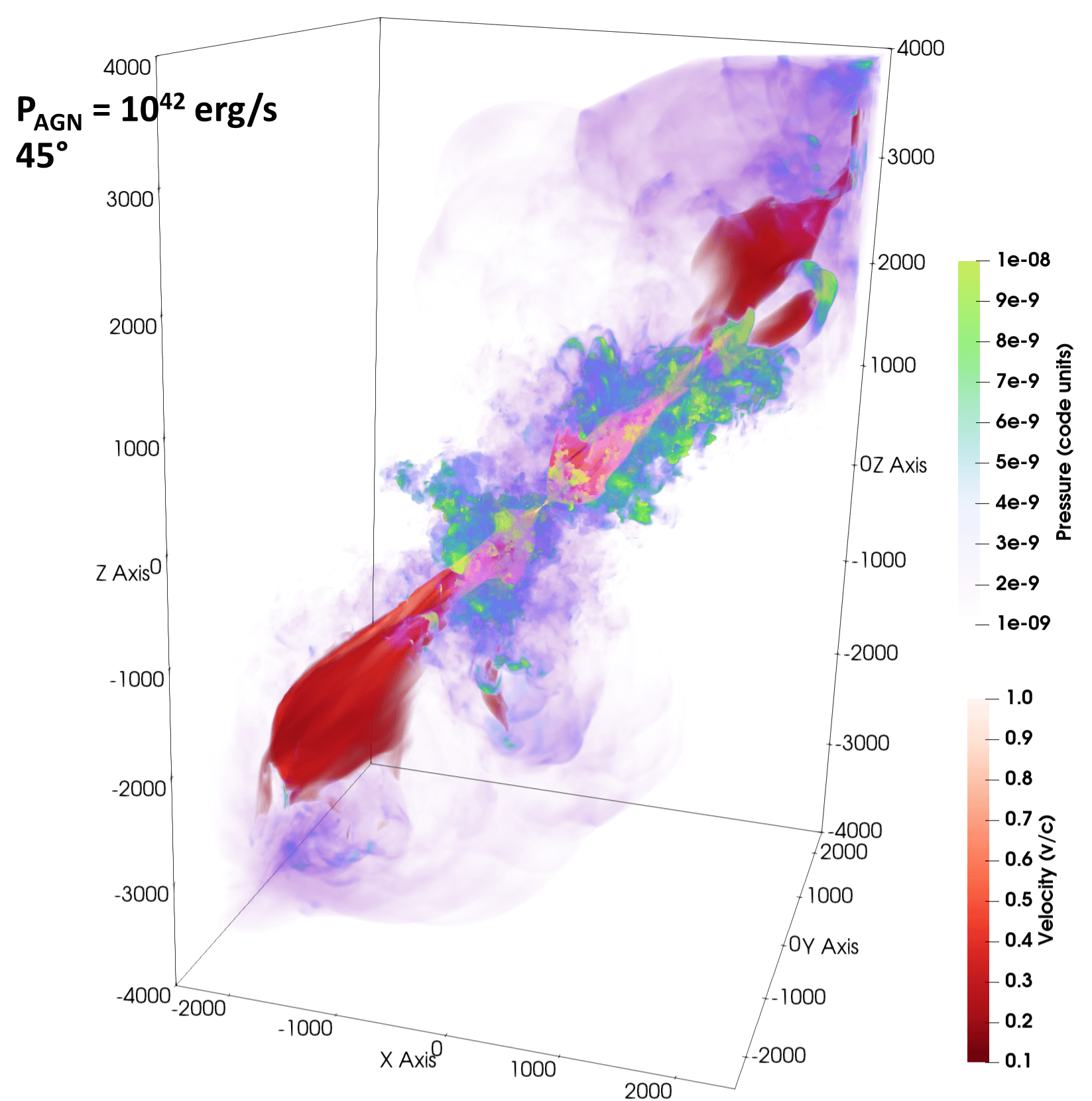}{0.33\textwidth}{}
          \fig{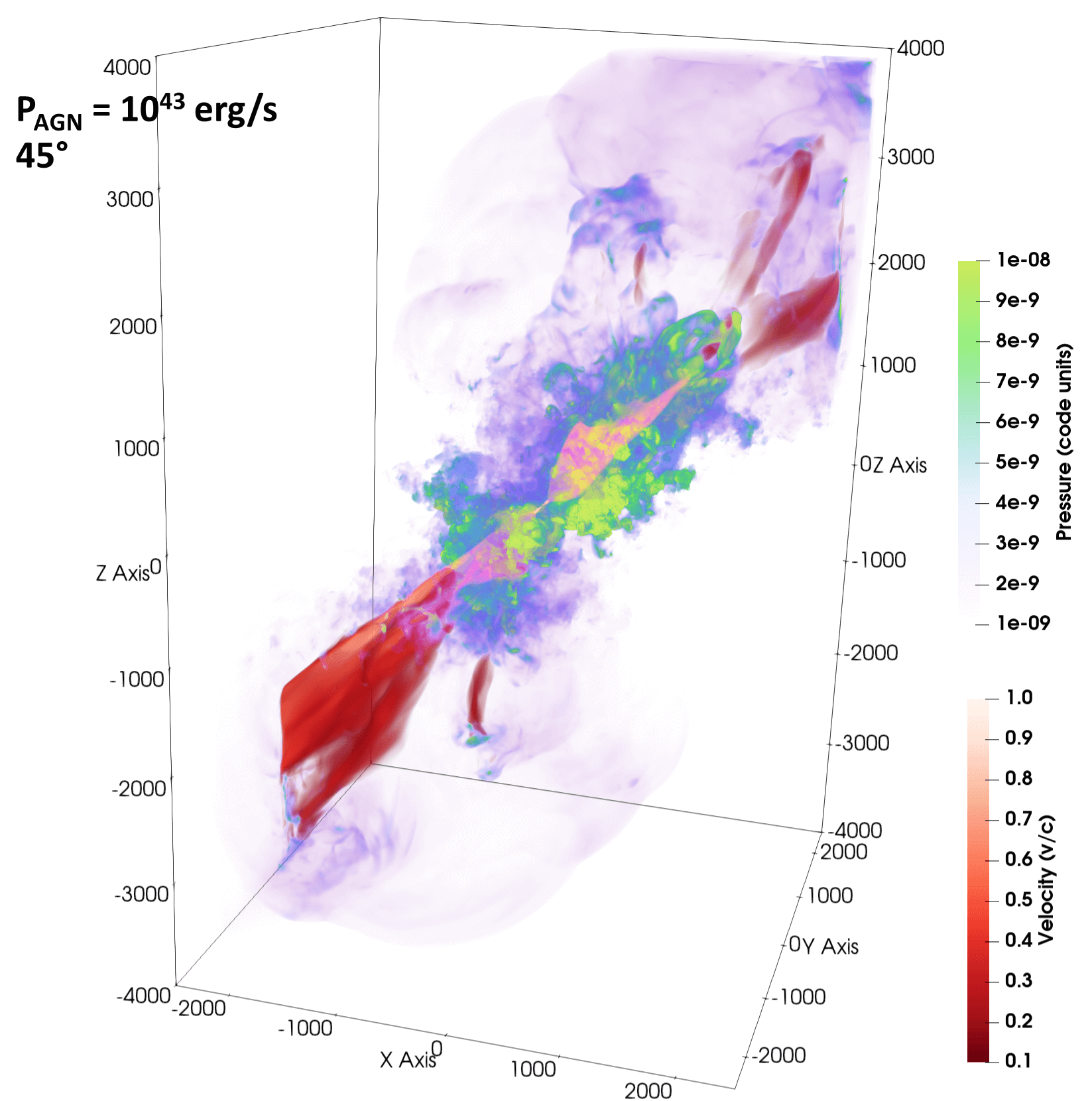}{0.33\textwidth}{}
          }
\gridline{\fig{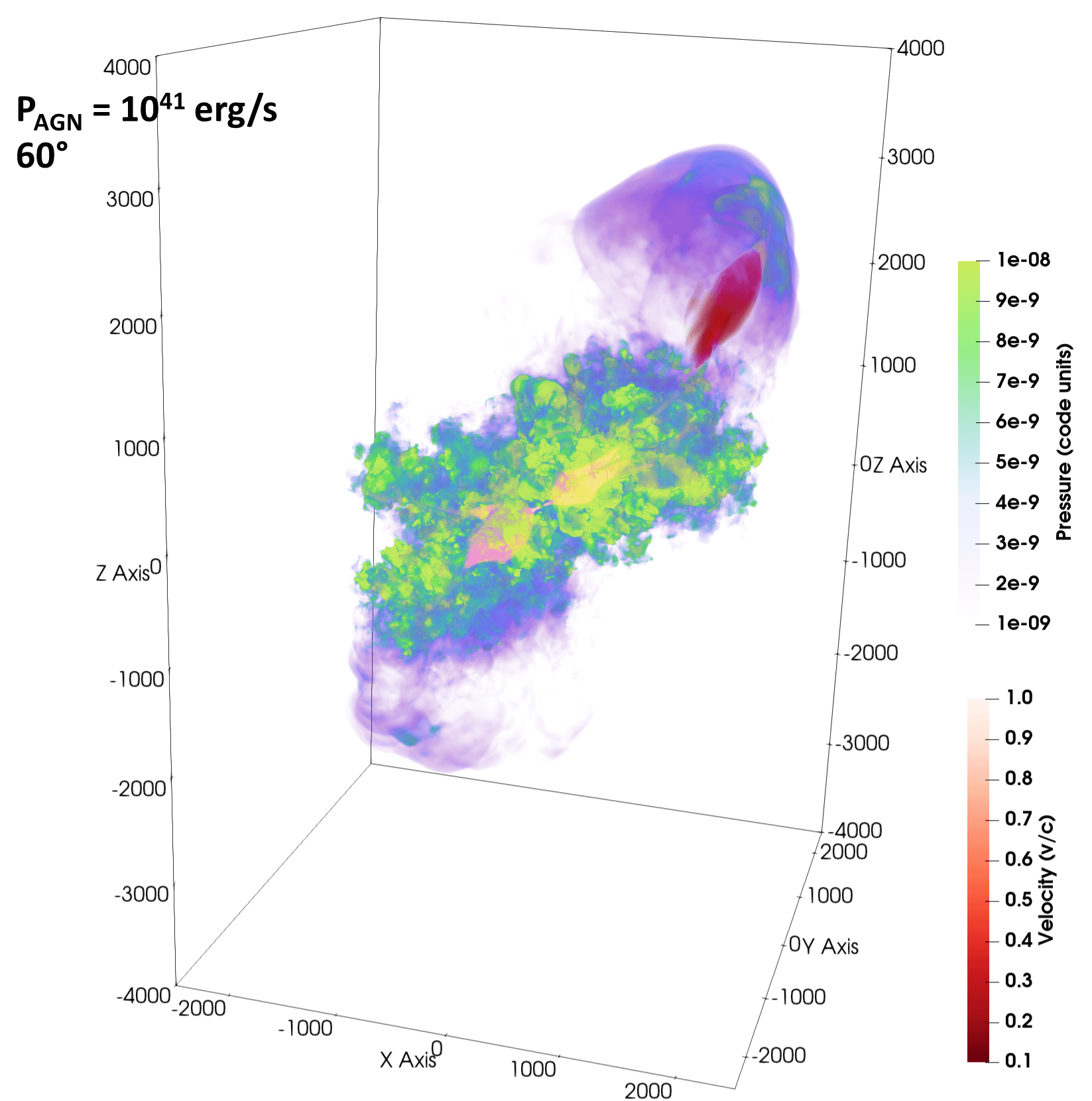}{0.33\textwidth}{}
          \fig{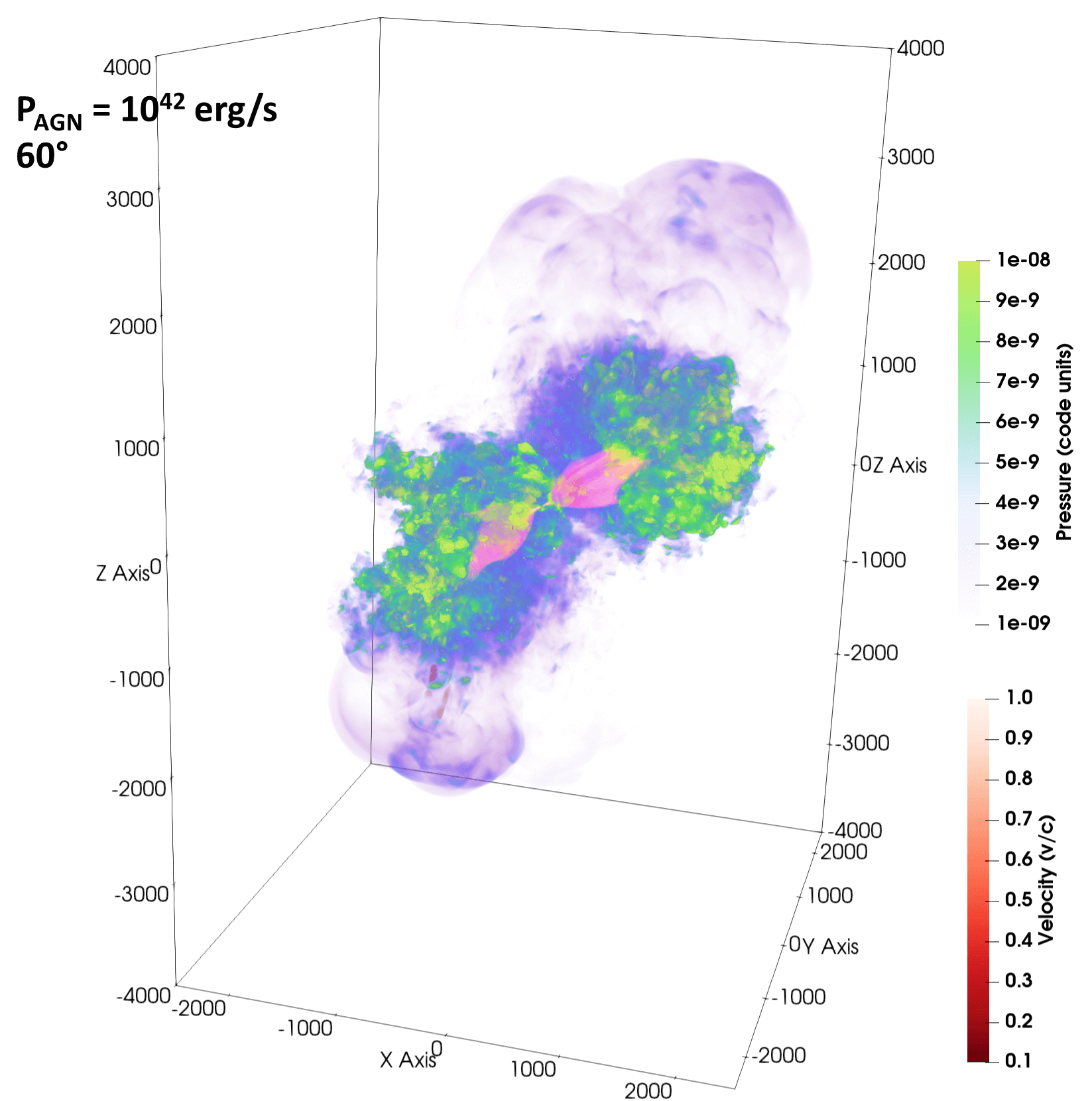}{0.33\textwidth}{}
          \fig{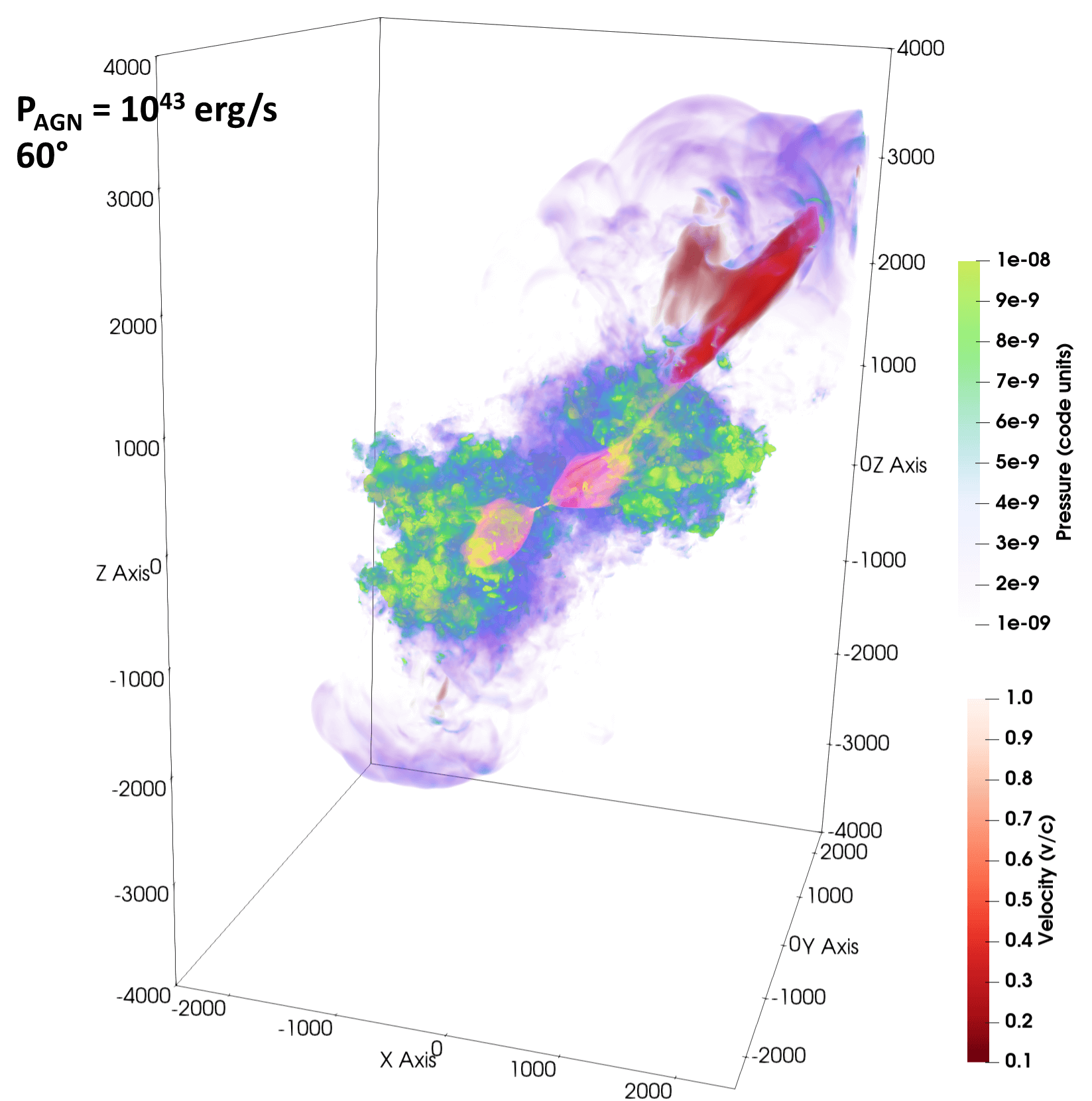}{0.33\textwidth}{}
          }
    \caption{Images show the velocity (reds) in units of c, and pressure (blue-green-yellow) in code units. The range in pressure was chosen to highlight the strongest shocks. Only velocities above $0.1$ c are shown. From left to right the AGN power is $10^{41}$ \ergs, $10^{42}$ \ergs, and $10^{43}$ \ergs. From top to bottom the inclination angles are $30^{\circ}$, $45^{\circ}$, and $60^{\circ}$. All simulations are shown at 600 kyrs.}
    \label{fig:vp1}
\end{figure}

After escaping the disk the shock front of the momentum-driven jet moves with a velocity of $30,000 - 60,000$ \kms ($0.1-0.2$ c). 
While the pressure driven bubble expands in all directions with a velocity of $3,000-5,000$ \kms. 
The exact speed of the shock fronts depends on how much of the dense ISM the jet interacted with. 
In figure \ref{fig:vp1} we plot how the momentum-driven jets are coincident with the shocked ISM gas showing areas of high velocity ($> 0.1$ c), and areas of high pressure. 
These are areas where we would expect the most radio and X-ray emission. 
Because our simulations do not include magnetic fields, and our grid resolution is too large to resolve the scale at which synchrotron radiation is generated, we can only make assumptions about where radio emission will be generated. 
In this case we assume that synchrotron radiation will be produced inside the high velocity ($> 0.1$ c) jets and inside the hot bubbles where the turbulent cascade will generate turbulence below the resolution of our simulations. 

In Figure \ref{fig:vp1} we show three simulations for three AGN angles ($30^{\circ}$, $45^{\circ}$, and $60^{\circ}$) and also three different AGN jet powers ($10^{41}$ \ergs, $10^{42}$ \ergs, and $10^{43}$ \ergs). 
In simulations where the jet interacts more with the ISM there is significantly more shocked gas than in simulations where the jet has a relatively free path. 
For the AGN angle of $30^{\circ}$ the jets on both sides are split and are beginning to dissipate into the turbulent bubble. 
The velocity and density difference will strongly contribute to Kelvin\textendash Helmholtz instabilities in the less dense jet gas. 
The resulting turbulence induces strong shocks in the ISM as shown in Figure \ref{fig:vp1}.

While there are noticeable differences as the AGN power increases, the strong similarities demonstrate that the structure of the ISM within the first $\sim 1$ kpc dominates the evolution of the jet and the shape of the outflow. 
The angle of impact of the jet on the closest dense clouds determines whether the jet will remain laminar or if Kelvin\textendash Helmholtz instabilities along with shocks will disrupt the jet before leaving the disk.

If there is a sufficiently dense cloud ($>100 $ cm$^{-3}$) of sufficient size ($> 50$ pc) and mass ($\sim 10^6$ \msun) located along the center line of the jet the the flow of the hot gas will split and the jet will become turbulent within a few kpc. 
This is the case with our simulations at $0^{\circ}$, $30^{\circ}$, and $60^{\circ}$ shown in Figure \ref{fig:vorticity}. 
If the cloud is not along the center line of the jet but still in the collimated flow then a portion of the jet will split off as can be seen in our simulations at $45^{\circ}$ and $75^{\circ}$ in Figure \ref{fig:vorticity}.

\begin{figure}
\gridline{
          \fig{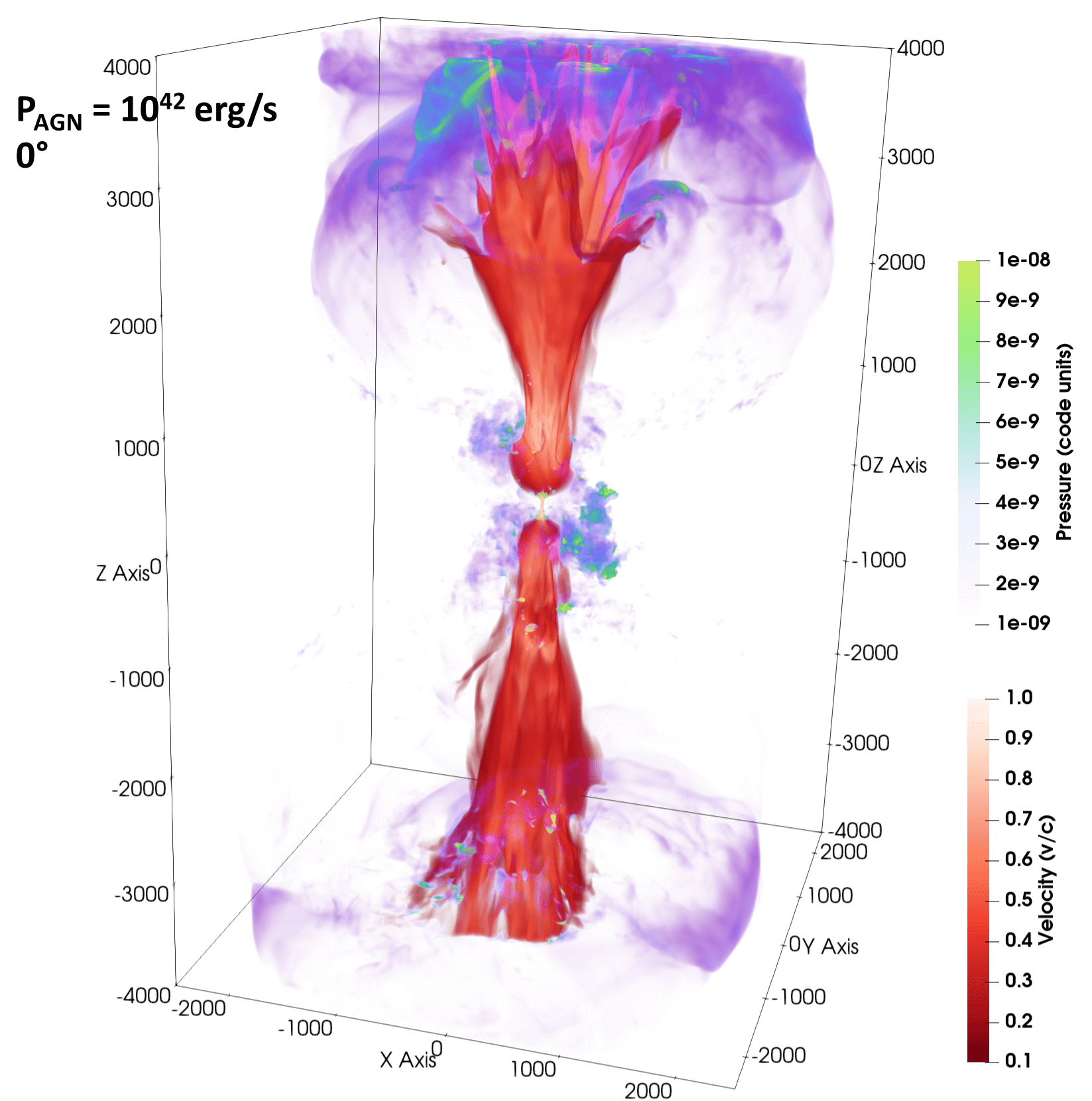}{0.33\textwidth}{}
          \fig{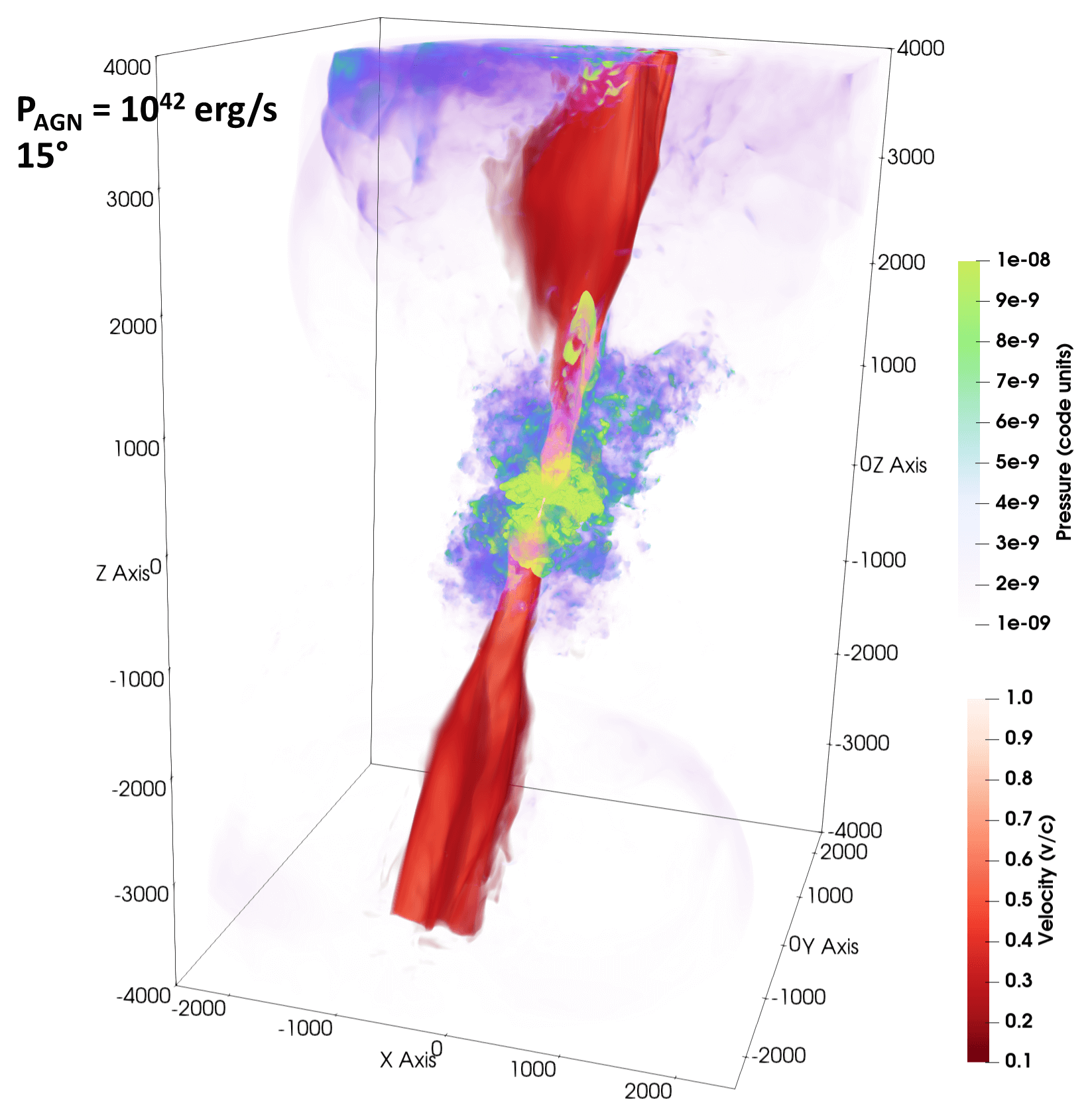}{0.33\textwidth}{}
          \fig{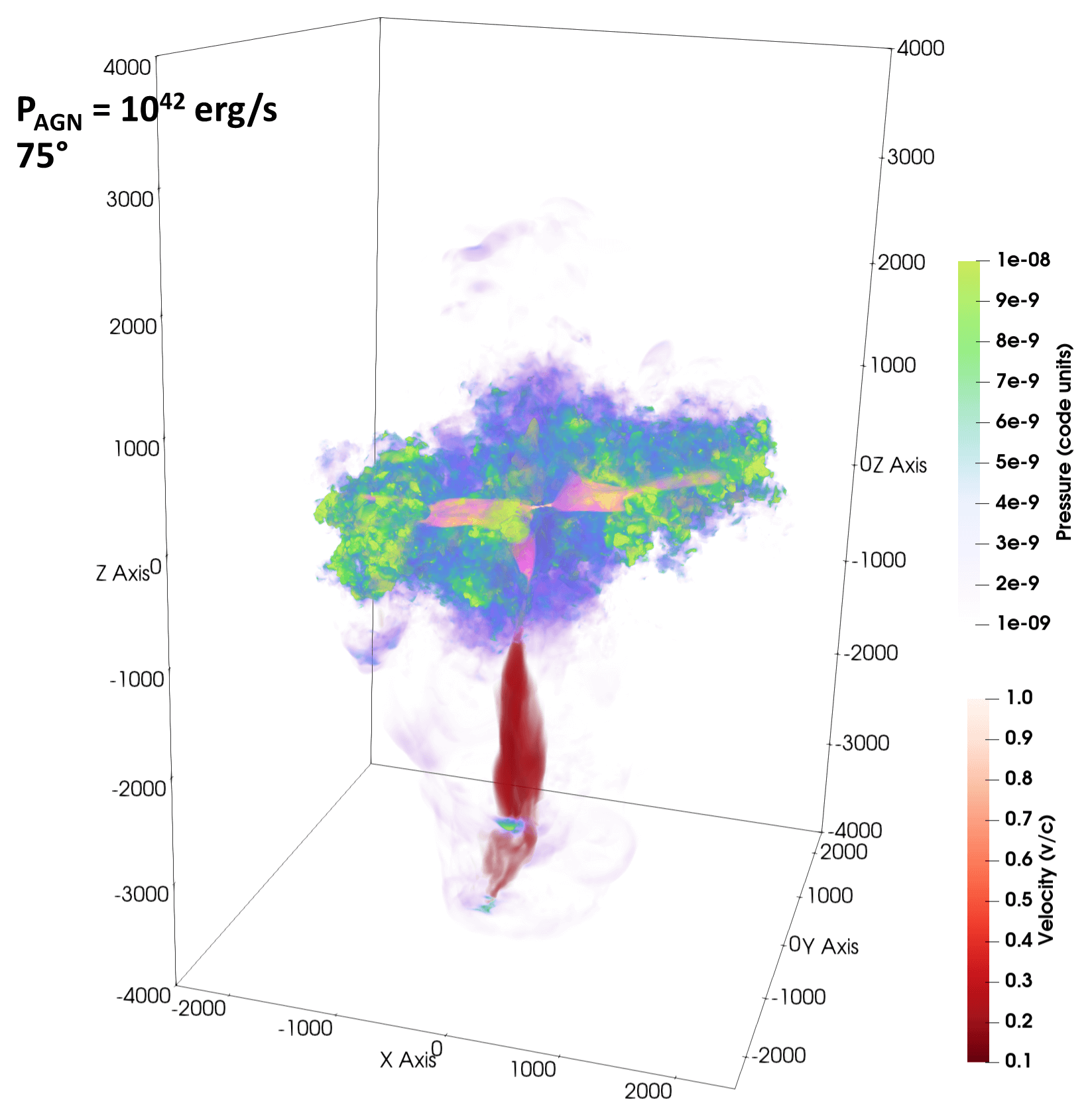}{0.33\textwidth}{}
          }
    \caption{Same as in Figure \ref{fig:vp1} with AGN power of $10^{42}$ \ergs, and from left to right inclination angles of $0^{\circ}$, $15^{\circ}$, $75^{\circ}$. All simulations are shown at 600 kyrs.}
    \label{fig:vp2}
\end{figure}

For all AGN inclination angles the jet has a relatively significant interaction with the ISM, with the exception of $15^{\circ}$ which had an unobstructed path on one side of the disk. 
As we show in Figures \ref{fig:vp1} and \ref{fig:vp2} only the simulation at $15^{\circ}$ shows a jet exiting the bottom side of the disk with little disruption to the jet. 
Thus with a clumpy ISM a jet is more likely to interact with multiple significant density gradients on its way out of the galactic disk. 
This will contribute to jet disruption and the growth of the turbulent bubble surrounding the jet.

The downward jet in our simulation with an AGN inclination angle of $15^{\circ}$ manages to reach the edge of the computational grid after 150 kyr while all other simulations that do reach the edge do so between 300 kyr and 500 kyr. 
Thus the ISM within the first kpc has a significant effect on the speed of the jet driven forward shock. 

\section{Discussion}\label{sec:discussion}

As mentioned in Section \ref{sec:setup} we are investigating how the ISM determines AGN driven galactic outflow morphology across six orders of magnitude of AGN jet power. 
The AGN powers considered straddle the traditional break point between FR I and FR II morphologies. 
Most of our simulations are the low power analogs of the high power AGNs in a clumpy medium modeled by \cite{2016MNRAS.461..967M, 2018MNRAS.479.5544M}. 
Because we used the same randomly generated ISM for all our simulations we could compare how the angle of inclination and the AGN power affected the morphology of the outflows versus a clumpy ISM structure. 
While the physical size of our simulations does not encompass the physical sizes of typical FR I and II radio lobes, our simulations do provide insight into the initial growth and structure of AGN driven galactic outflows in the first few kpc. 
We find that for low power AGNs the structure of the ISM can have a significant impact on whether a jet will form a coherent, laminar outflow, or if the jet will be significantly suppressed, disrupted, split, or deflected in some way. 
But the importance of the ISM structure is greatly reduced for higher power AGNs. 
Between $10^{44}-10^{45}$ \ergs the power input of the ISM becomes sufficient to fully disrupt the ISM regardless of the presence of massive ($\sim 10^{7}$ \msun) ISM structures in the path of the jet. 
Above this break point the AGN jets will always form coherent, laminar outflows. 
We do not consider the impact of cluster environment as it is outside the physical domain of our simulations. 
The various outflow morphologies seen in our simulations correspond to a variety of asymmetric or otherwise disrupted radio lobes observed in nearby galaxies.

\begin{figure}
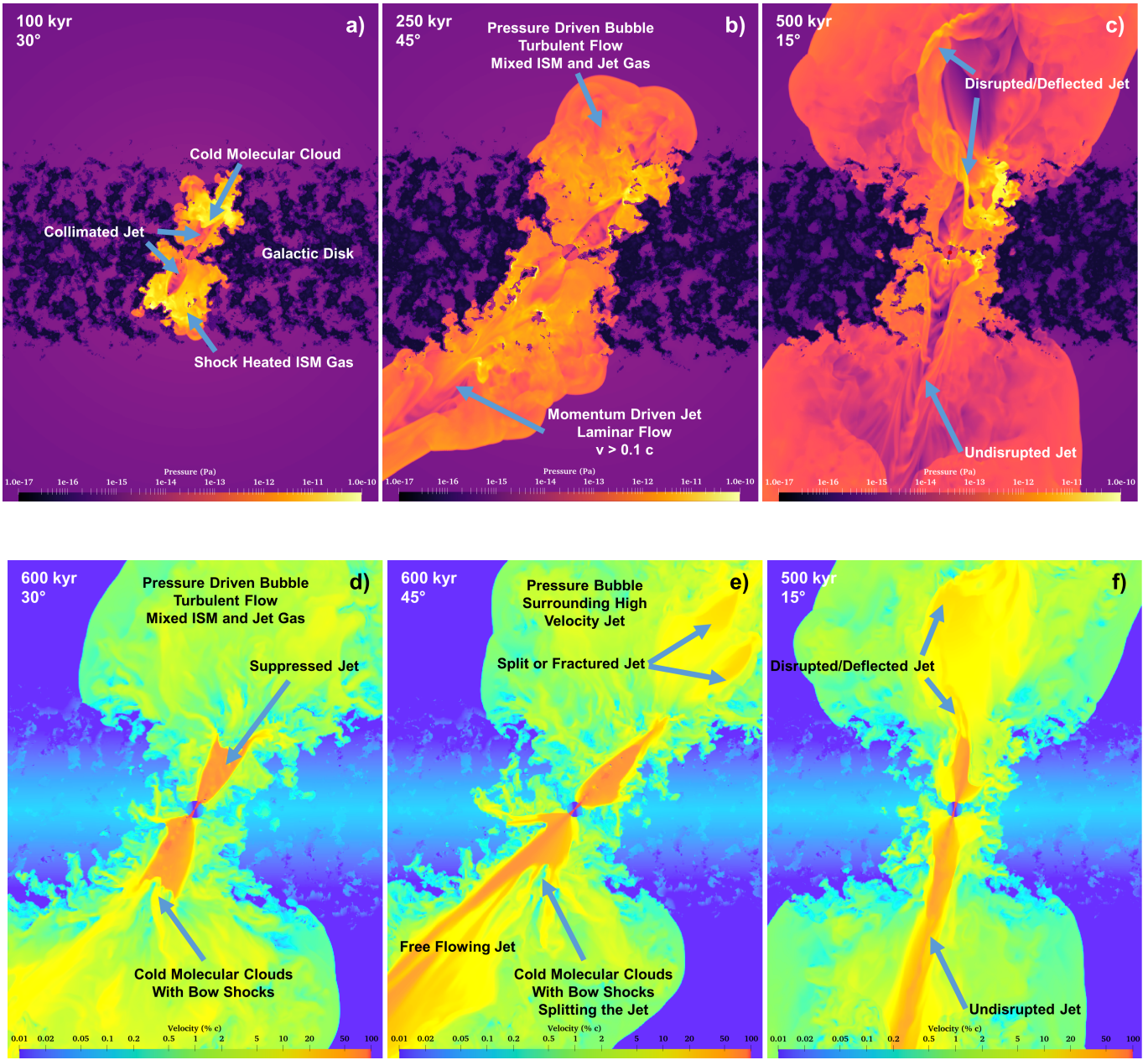

\gridline{\fig{discussion_a.png}{0.33\textwidth}{}
          \fig{discussion_b.png}{0.33\textwidth}{}
          \fig{discussion_c.png}{0.33\textwidth}{}
          }
\gridline{
          \fig{discussion_d.png}{0.33\textwidth}{}
          \fig{discussion_e.png}{0.33\textwidth}{}
          \fig{discussion_f.png}{0.33\textwidth}{}
          }
    \caption{Examples of different configurations of bubble and jet growth. In each image the central AGN region can be seen as circle with a biconical jet. This region is held constant in the simulations. In each case there is a collimated jet surrounded by mixed jet and ISM gas. This is similar to the jet propagation model of \cite{2011ApJ...740..100B}. In some cases large, cold molecular clouds block the path of the jet. These configurations are explained in the text.}
    \label{fig:cartoon}
\end{figure}

In Figure \ref{fig:cartoon} we show examples of the different morphology types found in our simulations of low power ($P_{AGN} = 10^{42}$ \ergs) AGN jets. 
For the first few hundred parsecs, after an initial expansion, the AGN jet forms a collimated flow similar to the jet propagation model of \cite{2011ApJ...740..100B}. 
A cocoon of shocked ISM gas surrounds the largely undisturbed laminar flow of the jet. 
At this point any radio emission would form a compact source object (CSO).

An unimpeded jet, or a jet through a uniform ISM, would form an elongated momentum-driven jet, as in Figure \ref{fig:cartoon} c), or f). 
There would be a narrow cocoon of mixed ISM and jet gas surrounding it which will expand at much slower velocities than the forward shock of the momentum-driven jet. 
An undisrupted jet, such as the one found in our simulation with an AGN inclination angle of $15^{\circ}$, will produce few shocks and a less prominent turbulent cloud surrounding the jet. 
As seen in Figure \ref{fig:vp2} (center image) there is very little shocked gas below the disk surrounding the jet. 
And in Figure \ref{fig:00xraytime} (top center) there is negligible soft X-ray emission compared to the other simulations. 
Thus the momentum driven jet is surrounded by a thin layer of hot x-ray emitting gas with a very small turbulent bubble around it.

Other possible configurations shown in Figure \ref{fig:cartoon} include a split jet (e), a disrupted/deflected jet (c) and (f), and a suppressed jet (d). 
If the dense clouds are not directly along the center line of the jet then you get the case shown in Figure \ref{fig:cartoon} e) where shock fronts will form around the clouds, but the jet will not be significantly disrupted. 
But because the clouds still interact with the jet the turbulent bubble will contain more mixed ISM and jet gas. 
The clouds may also deflect the jet, as in c), or may stop it all together as in d). 

Shells of hot X-ray emitting gas from the jet-ISM interaction have been observed in many galaxies with CSS/GPS sources, or LSOs \citep{2005xrrc.procE7.09K,2015ApJ...801...17L,2017ApJ...851...87O,2018A&A...619A..75M,2018ApJS..235...32S,2020MNRAS.497..482L,2020ApJ...894..157M,2021ApJS..252...31J}. \citet{2005xrrc.procE7.09K} using Chandra observed a hot shell of shocked gas surrounding the radio lobes in Centaurus A. 
This they concluded came from the jet-ISM interaction and is similar to the shocked shells in our simulations. 
In M 87 \citet{2005xrrc.procE7.09K} observed three expanding shells of hot gas which they interpreted as coming from three separate AGN events. 
The radio lobes in M 87 are coincident with the X-ray cavities created by the youngest of these events. 
They also noted a fragmentation, or bifurcation, in the shocked ISM material lifted up by the jet. 
These kinds of filamentary structures are consistent with the complex shock fronts and entrained ISM gas as seen in Figures \ref{fig:vp1} and \ref{fig:vp2}.

Optical spectroscopy of CSS radio sources have found Doppler shifted emission lines consistent with jet-ISM interactions of multiple components \citep{2002AJ....123.2333O,2018MNRAS.474.5319H,2019A&A...627A..53H,2019ApJ...879...75H,2020NewAR..8801539H,2021ApJ...908..221L}. 
The velocities observed (a few hundreds \kms~to $\sim$1,000 \kms) are consistent with the measured velocities of the cold components of our simulations (see Figure \ref{fig:masspdf}). 
In young jets only the hot and warm gas have had time to accelerate, while the cold gas takes time to accelerate up to terminal velocity \citep{2018A&A...617A.139S}. 

Finally there is a possibility of the significant deflection of a jet as seen in the simulation at $75^{\circ}$. 
In that case the jet was deflected downwards at an angle of $\sim 80^{\circ}$ with respect to the original jet direction. 
AGN jet deflections of this type have been observed \citep{2006A&A...455..161L,2010MNRAS.402...87A}. 
In the case of IC 2746 and ESO 428-14 \citep{2006A&A...455..161L} at $\approx2,500$ pc and $\approx500$ pc from the central black hole respectively there is a radio bright spot where the jet runs into a massive cloud in the ISM and the jet is redirected to nearly perpendicular to the original jet direction. 
In the case of Mrk 573 and Mrk 1014 the jets have not been deflected but have been stopped and confined in the galactic disk. 
We assume that whether a jet gets deflected, split, or stopped will depend on the total mass of the cloud and the geometry of the jet-cloud interaction.

An idealized outflow set up assumes a smooth ISM given by \citep{1972ApJ...174L.123K,2012ApJ...760...77A}
\begin{equation}
    \rho = \rho_0\left(\frac{a_0}{z}\right)^\beta
\end{equation}
where $\rho_0$ and $a_0$ are the central density and characteristic length scale respectively. 
The value of $\beta$ determines the steepness of the ISM. 
As shown in \cite{2012ApJ...760...77A} the linear extent of the radio lobes are a function of $\beta$ and $P_{AGN}$. 
In describing the evolution of CSOs to large-scale source objects (LSO) \cite{2012ApJ...760...77A} assumed a smooth ISM to derive the evolutionary tracks of CSOs. 
In their morphological classification and evolution of sources they assumed that $P_{AGN}$ is the determining factor in evolution of CSOs. 
For a given morphological type there was a one-to-one correspondence between $P_{AGN}$ and the evolutionary path. 

Alternatively, a more complex set of semi-analytic models \citep{2015ApJ...806...59T,2018MNRAS.475.2768H} do much better at modeling the complex evolution of jet driven radio lobes and reproducing the FR I and FR II dichotomy. 
In their models the jets form the familiar cocoon with jets that evolve in to FR I morphologies getting disrupted or flaring close to the host galaxy. 
The ultimate evolution of radio lobes in their models was strongly affected by the range of pressure profiles in the cluster environment of their sample of simulated galaxies. 
Our simulations cover a physical scale much smaller then the models considered in \citet{2015ApJ...806...59T,2018MNRAS.475.2768H}, but the external cluster environments of active galaxies will definitely determine the final morphology of the radio lobes. 

But a clumpy ISM greatly complicates this picture. 
As our AGN jets grow they encounter density changes of up to six orders of magnitude before reaching the ISM-CGM boundary. 
With randomly generated ISMs, AGNs with the same $P_{AGN}$ will follow widely different evolutionary paths. 
And, as explained above, a change in three orders of magnitude of $P_{AGN}$ did not significantly change the morphology or total mass of the outflow (see Figure \ref{fig:outmass}) when comparing jets with the same angle of inclination. 
This would indicate that $P_{AGN}$, at least in the low power regime, is a secondary factor for CSO evolution compared to the AGN angle with respect to the disk and the ISM structure within a few kpc.

\cite{2012ApJ...760...77A} explain the various evolutionary pathways taken by CSOs as they evolve to become medium symmetric objects (MSOs) and eventually large symmetric objects (LSO). 
One of the evolutionary scenarios shown in Figure 2 of \cite{2012ApJ...760...77A} involves the jet interacting with the ISM causing the final radio source to be dimmer and having a morphological type consistent with an FR I source.
To determine the probability of a CSO evolving into either a FR I or FR II radio galaxy we would have to consider the probability of a clear path being available to the jet. 
Whether or not the path is ``clear" would depend on the jet power as higher power jets could more easily clear smaller clouds. 
Also it would depend on the average ISM density and the maximum cloud size in the ISM which would determine the probability of each of the evolutionary pathways mentioned by \citet{2012ApJ...760...77A}. 
Considering the survey of radio sources by \citet{2019MNRAS.488.2701M}, jet-cloud or jet-ISM interactions could potentially dominate the evolution CSOs with low power AGNs into either FR I, FR II, hybrid, or anomalous morphologies. 
As our simulations show, for low power AGNs the ISM can result in a number of complex outflow morphologies with bent, split, disrupted, or suppressed jets. 
With the added effects of the cluster environment of the host galaxies the evolution of AGN driven outflows is obviously complex and multifaceted. 

With our small sample of simulations we obviously cannot address these issues, but hopefully our simulations can prompt more work into jet-cloud interactions, especially the case when the cloud is not fully within the jet. 
And also into the evolution of partially disrupted jets after leaving the disk as opposed to fully idealized jets as is standard for many jet simulations into the CGM of FR I and II sources \citep{2016MNRAS.461L..46T,2017MNRAS.470.4530W,2019MNRAS.483.2465M,2019A&A...621A.132M,2021arXiv210604100S}. 
Having a clumpy ISM should significantly affect the onset of either hydro or magnetic instabilities.

\section{Conclusions}\label{sec:conclusions}

Based on our simulations AGN driven outflows have two major components (see Figure \ref{fig:cartoon}),
\begin{enumerate}
    \item momentum-driven jets of high velocity, low density gas with laminar flow.
    \item Pressure driven bubbles of multi-phase, turbulent gas with a range of velocities.
\end{enumerate}
Below a critical AGN power of $P_{AGN} \approx 10^{44}$ \ergs the relative size and growth rate of each component depends strongly on the interaction of the AGN jet with the clumpy ISM on the kpc scale. 
For AGN with $P_{AGN} > 10^{44}$ \ergs the ISM plays a minor role in final morphology of the AGN driven outflow.
If a low power jet interacts with a dense cloud in the ISM it will either slow, disrupt, or split the laminar flow of the momentum-driven jet. 
This interaction will increase the growth and total mass of the pressure driven, turbulent bubble. 
An outflow that is almost purely momentum driven is inefficient at driving a multi-phase outflow with $>95\%$ of the gas mass being very hot ($> 10^8$ K). 
But a pressure driven bubble will drive a multi-phase out flow with roughly equal parts hot and warm gas, with a significant fraction of cold gas. 
Thus an AGN driven galactic outflow driving a multi-phase outflow requires some level of jet-cloud interactions.

The relative growth of the two outflow components depends on the angle of the AGN jet with respect to the galactic axis. 
An AGN with its jet perpendicular to the galactic disk will produce a strong momentum-driven jet and a weak pressure driven bubble. 
An AGN directed into the disk will only form pressure driven outflows.
Thus as the AGN angle goes from perpendicular to parallel to the disk the momentum-driven jet will become weaker, while the pressure driven bubble will become stronger. 
This shift will not greatly affect the total mass in the outflow. 
But AGN angles between $30^{\circ}$ and $60^{\circ}$ may have a total mass outflow 2-3 times higher than other angles. 
As the AGN angle increases the composition of the outflow will go from a single hot phase to a multi-phase outflow, with significantly more warm and cold gas (see Figure \ref{fig:outmass}). 
Thus the presence of neutral or slightly ionized gas in the outflow can be useful in determining the extent of the jet-ISM interaction.

This general trend can vary due to the interaction of the jet with the clumpy ISM. 
For example, if a sufficiently large cloud of dense gas sits along the central axis of the jet, then Kelvin-Helmholtz instabilities will disrupt the momentum-driven jet of high velocity gas within a few kpc. 
As the momentum-driven jet dissipates it will feed the growth of the turbulent bubble. 
In Figure \ref{fig:vorticity} we see how split ($0^{\circ}$, $75^{\circ}$) or blocked ($30^{\circ}$, $60^{\circ}$) laminar flows transition into lower velocity turbulent gas either close to the ISM-CGM boundary or within a few kpc of the boundary. 
The variety of outflow morphologies shows how having a clumpy ISM can cause the stability of outflows to be significantly different than idealized situations, such as \cite{2019MNRAS.488.4926I}.

These general trends hold across three orders of magnitude ($10^{41} - 10^{43}$ \ergs) of jet power.
For simulations with the same AGN inclination angle but with different powers there was no significant change in the total outflow mass or outflow composition based on temperature. Thus, for these low power AGN, the dominant variables in determining the morphology of the outflow were the the orientation of the AGN with respect to the disk and the clumpy structure of the ISM, and not the AGN power. 
But this was not true for our simulations with $P_{AGN} > 10^{44}$ \ergs.
Our results are in agreement with \cite{2019MNRAS.488.2701M} who found that luminosity does not reliably predict the morphology of the radio lobes (i.e. FR type I or II) for lower power AGNs like the ones we model. 
Our results indicate that the ISM within a few kpc of the AGN can significantly determine the morphology of the outflow.

With a randomly generated clumpy ISM, across the full range of possible AGN angles, only two momentum-driven jets managed to escape the disk while still maintaining relativistic laminar flow. 
Only these outflows could eventually form FR II radio lobes. 
None of our low power simulations produced symmetric high velocity, momentum-driven jets of equal strength and cohesion from both sides of the galactic disk. 
But almost all produced symmetric bubbles with a turbulent mix of ISM and jet gas. 
If the galactic outflows from our simulations were categorized by radio lobe morphology most would be classified as asymmetric, hybrids, or compact sources (so called ``FR 0s") \citep{2020A&A...642A.107C}.
From a sample of more than 5,800 radio bright sources \citep{2019MNRAS.488.2701M} only $\sim 7\%$ were FR II sources, with a similar number being hybrid sources. 
But the majority of sources in their survey were classified as small or compact sources.  
We recognize that we do not have a true stochastic sample, but this indicates an interesting avenue of investigation into how a random clumpy ISM determines the probability of different outflow morphology types, especially asymmetric or hybrid morphologies.

\acknowledgments
This research was supported by the appointment of Ryan Tanner to the NASA Postdoctoral Program (NPP) at the Goddard Space Flight Center. 
The NPP is administered by Universities Space Research Association (USRA) under contract with NASA. 
The simulations were run on the Discover cluster operated by the NASA Center for Climate Simulation at the Goddard Space Flight Center. 
The authors wishes to acknowledge the helpful comments from Edmond Hodges-Kluck and Anna Ogorzalek.
The authors also wish to acknowledge the helpful comments by the anonymous reviewer.

\bibliography{main}{}
\bibliographystyle{aasjournal}


\end{document}